\documentstyle[a4,amsmath,amssymb,latexsym,epsfig,citesort]{article}
\def \d{{\mathrm{d}}}

\def \pd{\partial}

\def \tl#1{\overset{\kern 1pt\circ}{#1}}
\def \TL#1{\overset{\kern -3pt \circ}{#1}}
\def \TLL#1{\overset{\kern -7pt \circ}{#1}}

\def \burger{{\boldsymbol{b}}}
\def \frank{\boldsymbol{\Omega}}

\begin{document}
\title{{\bf Nonsingular stress and strain fields of dislocations and disclinations in first strain 
gradient elasticity}}
\author{{\bf Markus Lazar\footnote{Corresponding author. 
{\it E-mail address:} lazar@lmm.jussieu.fr (M.~Lazar)}\;
and G{\'e}rard A.~Maugin}\\ \\
Laboratoire de Mod{\'e}lisation en M{\'e}canique,\\
        Universit{\'e} Pierre et Marie Curie,\\
	4 Place Jussieu, Case 162,\\	
	F-75252 Paris Cedex 05, France\\
}

\date{\today}    
\maketitle
\begin{abstract}
The aim of this paper is to study the elastic stress and strain fields of 
dislocations and disclinations 
in the framework of Mindlin's gradient elasticity.
We consider simple but rigorous versions of 
Mindlin's first gradient elasticity with one material length 
(gradient coefficient).
Using the stress function method, we find modified stress functions
for all six types of Volterra defects (dislocations and disclinations) 
situated in an isotropic and infinitely extended medium.
By means of these stress functions, we obtain exact analytical solutions 
for the stress and strain fields of dislocations and disclinations.
An advantage of these solutions for the elastic strain and stress 
is that they have no singularities at 
the defect line. 
They are finite and have maxima or minima 
in the defect core region. The stresses and strains are either zero
or have a finite maximum value at the defect line. The maximum value 
of stresses may serve as a measure of the critical stress level when 
fracture and failure may occur.
Thus, both the stress and elastic strain singularities are removed 
in such a simple gradient theory. 
In addition, we give 
the relation to the nonlocal stresses in Eringen's nonlocal elasticity
for the nonsingular stresses. \\

\end{abstract}
\vspace*{2mm}

\section{Introduction}
The traditional methods of classical elasticity break down at small distances from 
crystal defects and lead to singularities. 
This is unfortunate since the defect core
is an important region in the theory of defects.
Moreover, such singularities
are unphysical and an improved model of defects should eliminate them.
In addition, classical elasticity is a scale-free continuum theory 
in which no characteristic length appears.
Therefore, the classical elasticity cannot explain the phenomena near defects 
and at atomic scale. 

An extension of the classical elasticity is the so-called strain gradient
elasticity.
The physical motivation to introduce gradient theories
was originally given by Kr\"oner~\cite{Kroener63,KD66} in the early sixties.
The strain gradient theories extend the classical elasticity with
additional strain gradient terms. Due to the gradients,
they must contain additional material constants with the dimension of
a length, and hyperstresses appear. The hyperstress tensor 
is a higher order stress tensor given in terms of
strain gradients. 
In particular, the isotropic, higher-order gradient, linear elasticity was
developed essentially by Mindlin~\cite{Mindlin64,Mindlin65,ME68},
Green and Rivlin~\cite{GR64a,GR64b} and Toupin~\cite{Toupin64} 
(see also~\cite{Germain,Wu92,Maugin93}).
Strain gradient theories contain strain gradient terms and no
rotation vector and no proper couple-stresses  appear. In this way, they 
are different from theories with couple-stresses and Cosserat theory 
(micropolar elasticity). 
Only hyperstresses such as double or triple stresses
appear in strain gradient theories. 
Double stresses correspond to a force dipole and triple stresses
belong to a force quadrupole.
The next order would corresponds to a force octupole.

Stress and Hyperstress are physical quantities in a 
3-dimensional continuum mechanics. Within a framework of the 4-dimensional
spacetime continuum, stress and hyperstress translate into 
momentum and hypermomentum~\cite{hehl76,hehl97}. 

In the present work, 
we consider two simple but straightforward versions of 
a first strain gradient theory.
We investigate screw and edge dislocations and wedge and twist 
disclinations in the framework of incompatible strain gradient 
elasticity. 
We apply the ``modified'' stress function method 
to these types of straight dislocations and disclinations.
Using this method,
we derive exact analytical solutions for the stress and strain
fields demonstrating the elimination of ``classical'' 
singularities from the elastic field at the dislocation and disclination
line. Therefore, stresses and strains are finite within this 
gradient theory.
We obtain ``modified'' stress functions
for all types of straight dislocations and disclinations. 
In addition, we justify that these solutions are solutions 
in this special version
of Mindlin's first gradient elasticity. 
We also give the relation of nonsingular stresses of dislocations and disclinations
to Eringen's nonlocal elasticity theory. We show that these stresses
correspond to the ``nonlocal'' stresses. 

\section{Governing equations}
Following Mindlin~\cite{Mindlin64,Mindlin65,ME68} 
(see also~\cite{Wu92}), 
we start with the strain energy 
in gradient elasticity of an isotropic material.
We only consider first gradients of the elastic strain 
in this paper.  
In the small strain gradient theory the strain energy, $W$, is assumed to 
be a function
\begin{align}
\label{strain-en}
W=W(E_{ij},\pd_k E_{ij})
\end{align}
of the elastic strain, $E_{ij}$, and the gradient of the elastic strain,
$\pd_k E_{ij}$, which is sometimes called hyperstrain.
The $E_{ij}$ is dimensionless and the $\pd_k E_{ij}$ has
the dimension of the reciprocal of length. 
Obviously, no elastic rotation and gradients of it appear in~(\ref{strain-en}).  
The elastic strain can be a gradient of the displacement $u_i$
\begin{align}
\label{strain1}
E_{ij}=\pd_{(i}u_{j)}=\frac{1}{2}\big(\pd_i u_j+\pd_j u_i\big)
\end{align}
or can have the following form~\cite{deWit73b,EL88,Edelen96,Lazar02a,Lazar02b,Lazar03a}
\begin{align}
\label{strain2}
E_{ij}=\pd_{(i}u_{j)}+\tilde\beta_{(ij)},\qquad E_{ij}=E_{ji}
\end{align}
where $\tilde\beta_{(ij)}$ denotes the incompatible part of the elastic strain.
In addition, the negative incompatible strain may be identified as 
the plastic strain ($\beta^P_{(ij)}\equiv-\tilde\beta_{(ij)}$). 
The strain~(\ref{strain1}) is the elastic strain in a compatible situation. 
On the other hand, (\ref{strain2}) corresponds to the elastic strain for the incompatible
case (e.g. dislocations and disclinations). In the incompatible situation the 
$u_i$ and $\tilde\beta_{ij}$ are not ``good'' physical quantities because they are discontinuous functions.
Thus, they cannot be physical state quantities. 
Because the elastic strain is a physical state quantity, it must be a continuous function.
It is worth to note that the incompatible strain~(\ref{strain2}) is similar in 
form to Mindlin's so-called relative strain which contains a micro-strain
term. But in our case this part is identified as the incompatible elastic 
strain. Thus, we deal with a theory of 
incompatible strain gradient elasticity. 
 
The most general form of the strain energy for a linear, isotropic, 
gradient-dependent elastic material is given by~\cite{Mindlin64,Mindlin65}
\begin{align}
\label{SE-Mindlin}
W&=\frac{1}{2}\lambda E_{ii}E_{jj}+\mu E_{ij}E_{ij}
+c_1 \big(\pd_j E_{ij}\big) \big(\pd_k E_{ik}\big)
+c_2 \big(\pd_k E_{ii}\big) \big(\pd_j E_{jk}\big)\nonumber\\
&\quad
+c_3 \big(\pd_k E_{ii}\big) \big(\pd_k E_{jj}\big)
+c_4 \big(\pd_k E_{ij}\big) \big(\pd_k E_{ij}\big)
+c_5 \big(\pd_k E_{ij}\big) \big(\pd_i E_{jk}\big).
\end{align}
The constants $\lambda$ and $\mu$ are the Lam{\'e} constants and the five 
$c_n$ are the additional constants (gradient coefficients) which appear in 
Mindlin's strain gradient theory~\cite{Mindlin64,Mindlin65}.
Thus, five strain gradient terms~\footnote{We want to note that
Feynman\cite{Feynman} used five gradient terms of the metric tensor to obtain 
a linear theory of gravity. In that sense the theory of gravity can be considered
as a four-dimensional ``strain'' gradient theory in which the metric 
is the strain tensor and gravitation represents a ``metrical elasticity'' of
space.}  
appear in the isotropic case.

On the one hand,
the Cauchy stress tensor is defined as (Hooke's law)
\begin{align}
\label{HL}
\sigma_{ij}:=\frac{\pd W}{\pd E_{ij}}=
\lambda\delta_{ij} E_{kk}+2\mu E_{ij}
\end{align}
and has the symmetry $\sigma_{ij}=\sigma_{ji}$.
The elastic stress has the dimension of force, $F_i$, per unit area $\d A_j$.
In addition, the elastic energy is completely symmetric 
in the strain and stress tensor so that the condition for the strain follows 
\begin{align}
\label{HL-inv}
E_{ij}=\frac{\pd W}{\pd \sigma_{ij}}
=\frac{1}{2\mu}\, \Big(\sigma_{ij}-\frac{\nu}{1+\nu}\delta_{ij}\sigma_{kk}\Big),
\end{align}
with $\lambda=2\mu\nu/(1-2\nu)$.

\begin{figure}[t]\unitlength1cm
\centerline{
\begin{picture}(7,3)
\put(0.0,1.0){\epsfig{file=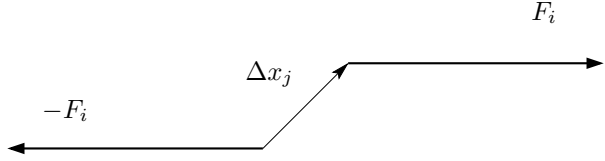,width=8cm}}
\put(3.2,2.0){$\Delta x_j$}
\put(0.5,1.5){$-F_i$}
\put(7.0,2.8){$F_i$}
\end{picture}
}
\caption{Double force stress $\tau_{ijk}$ acting on a plane with normal $n_k$.}
\label{fig:DF}
\end{figure}

\begin{figure}[t]\unitlength1cm
\centerline{
(a)
\begin{picture}(8,5)
\put(1.0,1.0){\epsfig{file=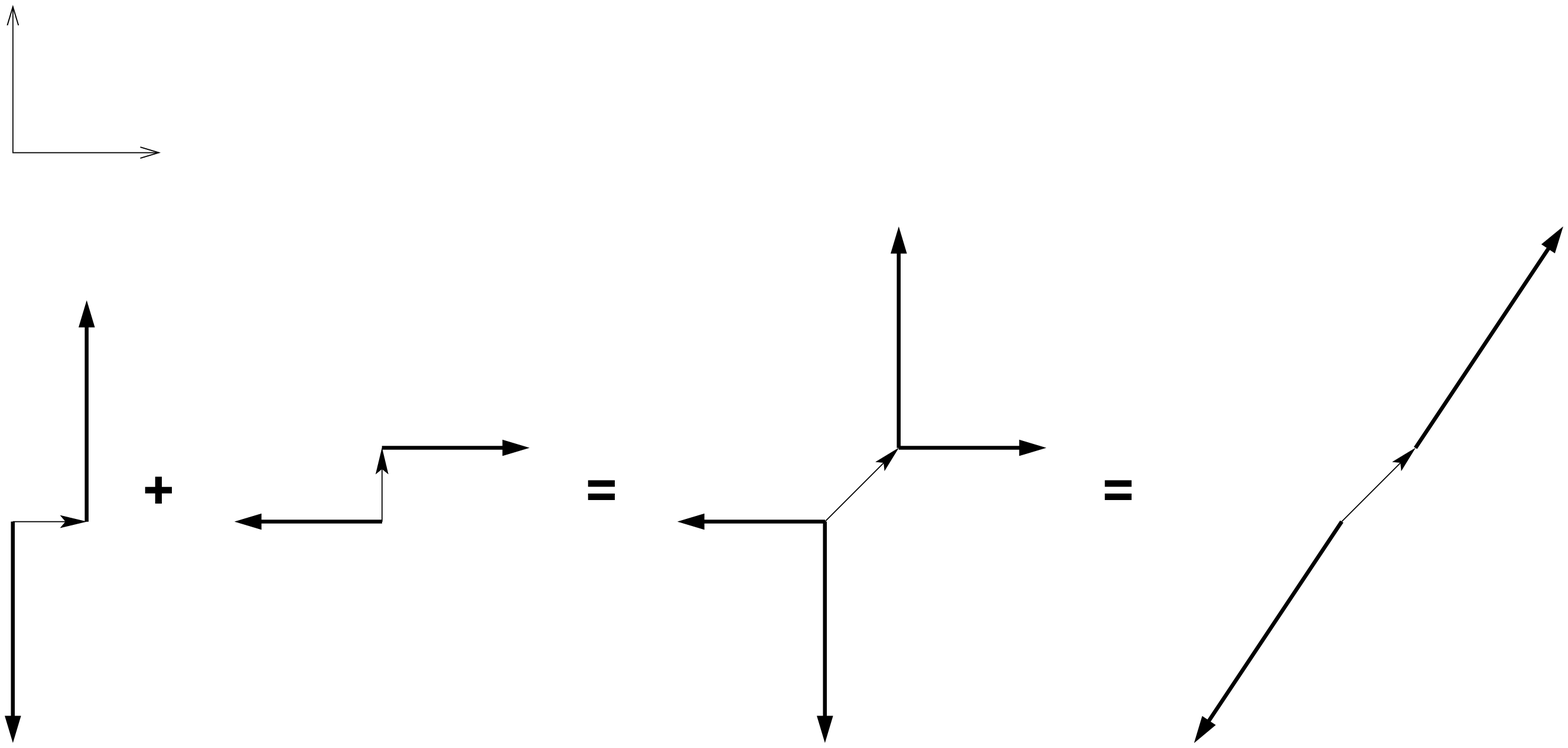,width=8cm}}
\put(2.5,0.5){$\tau_{xyk}$}
\put(1.0,0.5){$\tau_{yxk}$}
\put(5.0,0.5){$\tau_{(yx)k}$}
\put(1.6,3.7){$x$}
\put(0.7,4.5){$y$}
\end{picture}
}
\centerline{
(b)
\begin{picture}(8,4.0)
\put(1.0,1.0){\epsfig{file=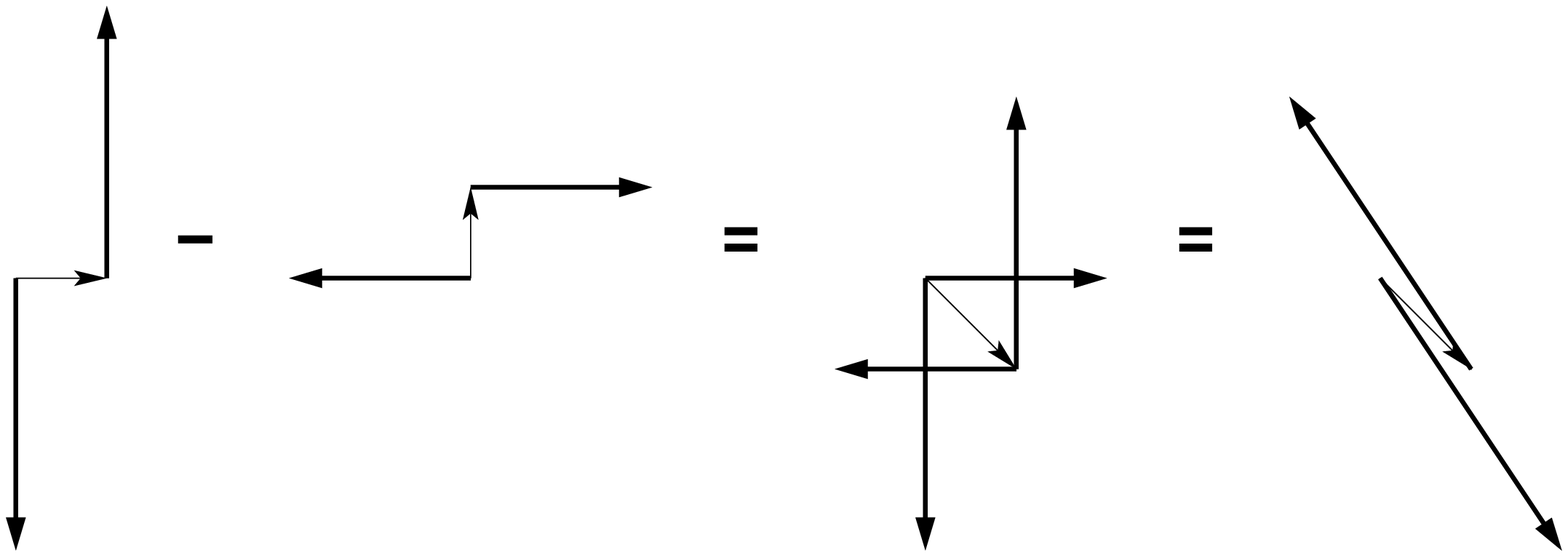,width=7.0cm}}
\put(2.5,0.5){$\tau_{xyk}$}
\put(1.0,0.5){$\tau_{yxk}$}
\put(5.0,0.5){$\tau_{[yx]k}$}
\end{picture}
}
\caption{Double force stresses (a) $\tau_{(yx)k}$ and (b) $\tau_{[yx]k}$ acting on a plane with normal $n_k$.}
\label{fig:DF2}
\end{figure}

On the other hand,
the hyperstress tensor is defined as follows (see also~\cite{Mindlin64,Mindlin65})
\begin{align}
\label{DS}
\tau_{ijk}:=\frac{\pd W}{\pd\big( \pd_k E_{ij}\big)}
          &=c_1\big(\delta_{ki}\pd_l E_{lj}+\delta_{kj}\pd_l E_{li}\big)
 	   +\frac{1}{2}\,c_2\big(\delta_{ki}\pd_j E_{ll}+\delta_{kj}\pd_i E_{ll}+2\delta_{ij}\pd_l E_{lk}\big)\nonumber\\
	   &\quad+2 c_3 \delta_{ij}\pd_k E_{ll}
	   +2 c_4 \pd_k E_{ij}
	   +c_5 \big(\pd_i E_{jk}+\pd_j E_{ik}\big).
\end{align}
It has the character of double forces per unit area.
The dipolar or double force acts through that area. 
In this case $\tau_{ijk}$ is a dipolar or double stress tensor. 
The first index of $\tau_{ijk}$ describes the orientation of the pair 
of forces $F_i$,
the second index gives the orientation of the lever arm $\Delta x_j$ between the forces 
and the third index denotes the orientation of
the normal $n_k$ of the plane on which the stress acts (see Fig.~\ref{fig:DF}). 
The double stress gives a contribution in the compatible as well as
the incompatible case.
In strain gradient theory it has the symmetry
\begin{align}
\tau_{ijk}\equiv\tau_{(ij)k}=\frac{1}{2}\big(\tau_{ijk}+\tau_{jik}\big),
\end{align}
and possesses 18 components. Thus, the symmetric part $\tau_{(ij)k}$ 
arises from
double forces without moment (see Fig.~\ref{fig:DF2}a). 
It can be resolved into the dilatational double stress $\tau_{llk}$ and
the traceless symmetric double stress $\tau_{(ij)k}-\frac{1}{3}\delta_{ij} \tau_{llk}$.
No antisymmetric double stresses $\tau_{[ij]k}$, which arise from double forces with 
moment (couple stresses), appear
\begin{align}
\tau_{[ij]k}=\frac{1}{2}\big(\tau_{ijk}-\tau_{jik}\big)=0,
\end{align}
which would correspond to rotation-gradients (see Fig.~\ref{fig:DF2}b). 
In general, a couple stress tensor has 9 components.

For vanishing external body forces, the force equilibrium equation can be derived from the principle of 
virtual work as (variation with respect to the displacement)~\cite{Mindlin64}
\begin{align}
\label{EC2}
\pd_j\big(\sigma_{ij}-\pd_k\tau_{ijk}\big)=0.
\end{align}
Since we consider an infinitely extended medium, we may neglect 
additional boundary conditions.
With Eqs.~(\ref{HL}) and (\ref{DS}), the force equilibrium~(\ref{EC2})
reads
\begin{align}
\label{EC2-2}
\pd_j\Big(\lambda\delta_{ij} E_{kk}+2\mu E_{ij}
-(c_1+c_5)\big(\pd_{i}\pd_l E_{lj}+\pd_{j}\pd_l E_{li}\big)
 	   -c_2\big(\pd_i\pd_j E_{ll}+\delta_{ij}\pd_k\pd_l E_{lk}\big)\nonumber\\
	   -2 \big(c_3\delta_{ij}\Delta E_{kk}+ c_4 \Delta E_{ij}\big)\Big)=0,
\end{align}
where $\Delta$ denotes the Laplacian. 
Such an expression~(\ref{EC2-2}) is quite formidable and hard to solve. 
Thus, a physical simplification could be useful.

First possibility of simplification: Let us follow a way in gradient elasticity similar to the way which 
Feynman~\cite{Feynman} went in linear gravity.
He pointed out that not all five gradient terms are necessary.
Because the gradient term to $c_5$ can be converted to the $c_1$ gradient term
by integration by parts. Therefore, the $c_5$ term may be omitted.
We set $c_5=0$.
In addition, the Cauchy stress tensor should fulfill 
an equilibrium condition of the kind as in elasticity
\begin{align}
\label{EC-C}
\pd_j\sigma_{ij}=0.
\end{align}
Consequently, in (\ref{EC2}) it must yield
\begin{align}
\label{EC-C2}
\pd_k\pd_j\tau_{ijk}=0,
\end{align}
and $\tau_{ijk}$ is self equilibrating.
Using (\ref{DS}) and $c_5=0$, Eq.~(\ref{EC-C2}) reads
\begin{align}
\label{EC-C3}
c_1\big(\pd_{i}\pd_j\pd_l E_{lj}+\Delta\pd_l E_{li}\big)
 	   +c_2\big(\pd_i\Delta E_{ll}+\pd_i\pd_k\pd_l E_{lk}\big)
	   +2 \big(c_3\pd_{i}\Delta E_{kk}+ c_4 \pd_j \Delta E_{ij}\big)=0.
\end{align}
It may be solved by 
\begin{align}
\label{GLS}
\pd_i\pd_j\pd_l E_{jl}\big(c_1+c_2\big)&=0,\nonumber\\
\Delta\pd_l E_{il}\big(c_1+2c_4\big)&=0,\nonumber\\
\Delta \pd_iE_{ll}\big(c_2+2c_3\big)&=0.
\end{align}
If we choose a scale such  that $c_4=\ell/2$, then
the solution of~(\ref{GLS}) is given by
\begin{align}
\label{EC}
c_1=-\ell,\qquad c_2=\ell,\qquad 
c_3=-\frac{\ell}{2},\qquad c_4=\frac{\ell}{2}.
\end{align}
The coefficient $\ell$ has the dimension of a force.
Thus, the coefficient $\ell$ may be chosen
\begin{align}
\ell=\frac{2\mu}{\kappa^2}
\end{align}
with only one ``gradient-coefficient'' or ``parameter of nonlocality''
\begin{align}
\kappa^{-2}> 0.
\end{align}
$\kappa^2$ has the dimension of $1/[{\text{length}}^2]$.
It is a positive material constant.
What we have recovered is nothing but the so-called Einstein choice
in three dimensions
scaled by a factor $\ell$ (see, e.g.,~\cite{KV92,Lazar02a}).
So, we can write down the expression
\begin{align}
\label{Einstein}
\frac{1}{\ell}\, \pd_k \tau_{ijk}
&=\left(\text{inc}\, \boldsymbol{E}\right)_{ij}
\equiv-\epsilon_{ikl}\epsilon_{jmn}\pd_k \pd_{m} E_{l n}\nonumber\\
&=\Delta E_{ij}-(\pd_j\pd_k E_{ik}+\pd_i\pd_k E_{kj})
+\delta_{ij}\pd_k\pd_l E_{kl}+\pd_i\pd_j E_{kk}
-\delta_{ij}\Delta E_{kk}
\end{align}
which is equivalent to the three dimensional linear Einstein tensor.
The linear Einstein tensor  
is the incompatibility of the strain (see~\cite{Kroener58,Kroener81}).
Thus, the three dimensional Einstein tensor is proportional 
to the divergence of the double force stress tensor.
The Einstein choice has been used in the gauge theory of dislocations
by Lazar~\cite{Lazar00,Lazar02a,Lazar02b} and by Malyshev~\cite{Malyshev00}.
Eq.~(\ref{Einstein}) leads to physical solutions for the screw 
dislocation~\cite{Lazar02a,Lazar02b,Malyshev00}.
The solution of an edge dislocation~\cite{Malyshev00} has a modified far field of the 
stress.
In fact, it does not coincide with the far field of the classical solution.
One way out is to modify the strain energy with additional
bend-twist terms~\cite{Lazar03a}. 
But then we are leaving the framework of strain gradient elasticity
which is not our aim.
Anyway, it is interesting to note that the three dimensional Einstein
choice which is used in the gauge theory of dislocations is
contained in the strain energy expression~(\ref{SE-Mindlin}) given by Mindlin. 

Second possibility of simplification:
If we require that the stress-strain symmetry of the elastic energy 
should also be valid for the gradient terms, and we think this
is quite natural, we have to use 
the following choice of the five gradient coefficients
\begin{align}
\label{choice}
c_1=0,\quad c_2=0,\quad c_3=\frac{\lambda}{2\kappa^2},\quad 
c_4=\frac{\mu}{\kappa^{2}},\quad c_5=0.
\end{align}
Then the strain energy can be rewritten in the following simple form
\begin{align}
\label{en-sy}
W=\frac{1}{2}\,\sigma_{ij}E_{ij}
+\frac{1}{2\kappa^2}\,\big(\pd_k\sigma_{ij}\big)\big(\pd_k E_{ij}\big).
\end{align}
One can see that $\kappa$ has the dimension of an inverse length.
Thus, by using the special choice~(\ref{choice}), we have obtained a simple
but still rigorous  version
of Mindlin's gradient theory which is a simple strain gradient as well 
as stress gradient theory. In particular, the elastic energy~(\ref{en-sy}) 
is symmetric with respect to the strain and the stress 
and also with respect to the strain gradient and the stress gradient
\begin{align}
\frac{\pd W}{\pd (\pd_k E_{ij})}=\frac{1}{\kappa^{2}}\, \pd_k \sigma_{ij},\qquad
\frac{\pd W}{\pd (\pd_k \sigma_{ij})}=\frac{1}{\kappa^{2}}\, \pd_k E_{ij}.
\end{align}
Then the double stress tensor has the following form
\begin{align}
\label{tau2}
\tau_{ijk}=\frac{1}{\kappa^{2}}\,\pd_k \sigma_{ij}
	 =\frac{1}{\kappa^{2}}\,\big(2\mu \pd_k E_{ij}+\lambda\delta_{ij} \pd _k E_{ll}\big).
\end{align}
Consequently, the double stress~(\ref{tau2}) is a simple gradient of the Cauchy stress.
Thus, it is a higher-order stress tensor.
Only $\kappa^{-2}$ is a non-standard coefficient of the theory.
Recently, double stresses similar in their form to~(\ref{tau2}) have also
been used in the mode-I crack problem~\cite{Ex98} and the mode-III crack 
problem~\cite{Fann,Georg}.

Let us mention 
that a higher order gradient elastic energy with the required symmetry
would have the following form
\begin{align}
\label{en-sy2}
W&=\frac{1}{2}\,\sigma_{ij}E_{ij}+
\frac{1}{2\kappa^2}\,\big(\pd_k\sigma_{ij}\big)\big(\pd_k E_{ij}\big)+
\frac{1}{2\hat\kappa^4}\,\big(\pd_l\pd_k\sigma_{ij}\big)\big(\pd_l\pd_k E_{ij}\big)\nonumber\\
&\quad
+\frac{1}{2\tilde\kappa^6}\,\big(\pd_m\pd_l\pd_k\sigma_{ij}\big)\big(\pd_m\pd_l\pd_k E_{ij}\big)
+\cdots,
\end{align}
where $\hat\kappa$ and $\tilde\kappa$ are higher order gradient coefficients.
It is interesting to note that the last term in Eq.~(\ref{en-sy2}) with the third strain gradient 
corresponds to a force octupole. 
But in the following we only consider the first gradient theory given 
by~(\ref{en-sy}).

In addition, we note that Altan and Aifantis~\cite{AA97} have used 
a similar choice like~(\ref{choice}) 
for the study of cracks within their version of strain gradient elasticity. 
Anyway, they did not use the notion of double stress. 
For that reason, they identified the Cauchy stress with the total stress
tensor. The price they had to pay was that the singularities are still present
in the stresses.
In order to regularize both the elastic strain and the stress singularities
Ru and Aifantis~\cite{RA} (see also~\cite{Aifantis03}) 
introduced a constitutive relation with gradients of the elastic strain 
and the stress multiplied by two different gradient coefficients. 
In our framework, which we use in this paper, it is not 
necessary to introduce such a constitutive relation because the Cauchy stress
is defined as the derivative of the strain energy 
with respect to the elastic strain and nothing else. 
On the other hand, Gutkin and Aifantis~\cite{GA99b} used another choice so 
that no double stresses appear. But, on the other hand, 
triple stresses which correspond to second gradient strain  must appear 
in their approach and, therefore, in the equilibrium equations.
But they neglected the triple stresses which is not straightforward.    
The strain gradient elasticity which contains two different gradient
coefficients proposed by Ru and Aifantis~\cite{RA93} 
is used by Gutkin and Aifantis~\cite{GA99,GA00,Gutkin00}
for dislocations and disclinations. 


By substituting (\ref{tau2}) into (\ref{EC2}), we obtain
\begin{align}
\label{EC3}
\big(1-\kappa^{-2}\Delta\big)\pd_j\sigma_{ij}=0.
\end{align} 
If we define 
\begin{align}
\label{CR1}
\tl\sigma_{ij}=\sigma_{ij}-\pd_k\tau_{ijk}	
	      =\lambda\delta_{ij} E_{ll}+2\mu E_{ij}-\kappa^{-2}\big(\lambda\delta_{ij}\Delta E_{ll}+2\mu \Delta E_{ij}\big),
\end{align}
Eq.~(\ref{EC3}) takes the form
\begin{align}
\label{EC1}
\pd_j\tl\sigma_{ij}=0.
\end{align}
The stress tensor $\tl\sigma_{ij}$ may be called total stress tensor, 
which is a kind of ``balanced stress''.  
In the gauge theory of defects it is identified as the background stress tensor. 
One other difference with Gutkin and Aifantis' approach~\cite{GA99b} is that
they consider (\ref{CR1}) as a modified constitutive relation while we 
obtained (\ref{CR1}) as field equation instead.

On the other hand, in the incompatible situation we have the additional 
field $\tilde\beta_{ij}$. Which equation corresponds to this field? 
One should obtain Eq.~(\ref{CR1}) as a variation of strain energy 
with respect to $\tilde\beta_{ij}$. But so far this does not work. 
The way out is to add an additional strain energy part to~(\ref{en-sy}), 
\begin{align}
W'=-\tl\sigma_{ij}E_{ij},
\end{align}
which in the compatible situation is
a null Lagrangian such that it gives no contribution in the variation
with respect to the displacement. 
The nontrivial traction boundary problems 
in the variational formulation of field theory
can be formulated by means of a null Lagrangian.
This procedure is well-known in the gauge theory of
dislocations~\cite{EL88,Edelen96,Lazar02a,Lazar02b,Lazar03a,Malyshev00}. 
When the null Lagrangian is added to the elastic Lagrangian (strain energy) it does not change the
Euler-Lagrange equation (force equilibrium) because the 
associated Euler-Lagrange equation, $\pd_j\tl\sigma_{ij}=0$, must be
identically satisfied.
Then for the incompatible case, Eq.~(\ref{CR1}) may be considered as the field equation 
corresponding to $\tilde\beta_{ij}$.
It is interesting that (\ref{CR1}) appears in the compatible as well as the incompatible
situation. Only the interpretation of (\ref{CR1}) is different.

Now we rewrite Eq.~(\ref{CR1}) and obtain 
an inhomogeneous Helmholtz equation 
for every component of the
Cauchy stress 
\begin{align}
\label{stress-fe}
\big(1-\kappa^{-2}\Delta\big)\sigma_{ij}=\tl\sigma {}_{ij}.
\end{align}
Since the factor $\kappa^{-1}$ has the physical dimension of a length, 
it defines an internal characteristic length in quite a natural way.
It is worth to note that Eq.~(\ref{stress-fe}) agrees with the field equation
for the nonlocal stress in Eringen's nonlocal elasticity~\cite{Eringen83,Eringen85,Eringen90,Eringen02}
and for the stress in the gradient elasticity given by Gutkin and Aifantis~\cite{GA99,GA00,Gutkin00}. 
If we consider dislocations and disclinations which have an axial symmetry,
Eq.~(\ref{stress-fe}) may be rewritten as a convolution integral
\begin{align}
\label{stress-nl}
\sigma_{ij}(r)=\int_V \alpha(r-r')\,\tl\sigma {}_{ij}(r')\, \d v(r'),
\end{align}
with the corresponding two-dimensional Green's function
\begin{align}
\label{green}
\alpha(r-r')& =\frac{\kappa^2}{2\pi}\,K_0\big(\kappa \sqrt{(x-x')^2+(y-y')^2}\big).
\end{align}
Here $K_n$ is the modified Bessel function of the second kind and 
$n=0,1,\ldots$ denotes the order of this function.
In comparison with Eringen's nonlocal theory of elasticity
the Green function~(\ref{green}) may be identified as a
nonlocal kernel introduced by Ari and Eringen~\cite{AE83}.
Using the inverse of Hooke's law for the stress $\sigma_{ij}$ and $\tl\sigma {}_{ij}$, 
it follows that the elastic strain can be determined from the equation
\begin{align}
\label{strain-fe}
\big(1-\kappa^{-2}\Delta\big)E_{ij}=\tl E {}_{ij},
\end{align}
where $\tl E {}_{ij}$ is the classical strain 
tensor.
Because the strain tensor fulfills an inhomogeneous Helmholtz equation,
we may rewrite~(\ref{strain-fe}) as a nonlocal relation for the strain
\begin{align}
\label{strain-nl}
E_{ij}(r)=\int_V \alpha(r-r')\,\tl E {}_{ij}(r')\, \d v(r').
\end{align}
Of course, such a relation (\ref{strain-nl}) does not appear 
in Eringen's theory~\cite{Eringen02} of nonlocal elasticity.
In his theory, the displacement and the elastic strain are the
same as in classical elasticity.
Using Hooke's law, we can combine (\ref{stress-fe}) and (\ref{strain-fe})
to a gradient like relation
\begin{align}
\label{HL-gr}
\big(1-\kappa^{-2}\Delta\big)\sigma_{ij}=
\big(1-\kappa^{-2}\Delta\big)\big(\lambda\delta_{ij} E_{kk}+2\mu E_{ij}\big).
\end{align}
It is interesting to note that Eq.~(\ref{HL-gr}) has the same form as
the gradient constitutive relation given in~\cite{RA93} 
if their two different gradient coefficients are the same gradient
coefficients.
For modified solutions of the stress and elastic strain fields
we require that the far field of them should agree with the classical 
expressions and they should be free from the classical singularities 
at the defect line. Therefore, we are looking for nonsingular 
solutions for both the stress and the elastic strain. 
It depends on the taste of the reader to consider alternatively 
such constraints
as physically motivated boundary conditions. 

Using the decomposition~(\ref{strain2}), we obtain the
coupled partial differential equation
\begin{align}
\label{PDE-coup}
\big(1-\kappa^{-2}\Delta\big)\big[\pd_{(i}u_{j)}-\beta^P_{(ij)}\big]=
\pd_{(i}\tl u {}_{j)}-\tl \beta {}^P_{(ij)},
\end{align}
where $\tl u {}_i$ denotes the  
displacement field and $\tl \beta {}^P_{ij}$ is the 
plastic distortion in classical defect theory (see, e.g.,~\cite{deWit73b}).
Thus, if the following equations are fulfilled
\begin{align}
\label{dist-HE}
&\big(1-\kappa^{-2}\Delta\big)\beta_{ij}=\tl \beta {}_{ij},\\
\label{plast-HE}
&\big(1-\kappa^{-2}\Delta\big)\beta^P_{ij}=\tl \beta {}^P_{ij},
\end{align} 
the equation for the displacement field~\footnote{If $\beta^P_{ij}=0$ (compatible distortion),
the inhomogeneous Helmholtz equation, which was already proposed by Aifantis~\cite{Aifan92},
Ru and Aifantis~\cite{RA},  
$\big(1-\kappa^{-2}\Delta\big)u_{i}=\tl u {}_{i}$
is obtained without further assumptions.},
\begin{align}
\label{u-HE}
\big(1-\kappa^{-2}\Delta\big) u_{i}=\tl u {}_{i},
\end{align}
is valid for the incompatible case. 
Thus, for defects (dislocations, disclinations) 
the inhomogeneous parts of Eqs.~(\ref{plast-HE}) and (\ref{u-HE})
are fields with discontinuities (jumps).
We note that Eq.~(\ref{u-HE}) was used by Gutkin and Aifantis~\cite{GA96,GA97,GA99}
in order to calculate the displacement fields for screw and edge dislocations.

In nonlocal elasticity
Eringen~\cite{Eringen83,Eringen85,Eringen90,Eringen02} found the 
two-dimensional kernel~(\ref{green}) 
by giving the best match with the Born-K{\'a}rm{\'a}n model of the 
atomic lattice dynamics and the phonon dispersion curves.
The length, $\kappa^{-1}$, may be
selected to be proportional to the lattice parameter $a$ 
for a single crystal, i.e.
\begin{align}
\kappa^{-1}=e_0\, a,
\end{align}
where $e_0$ is a non-dimensional constant~\cite{Eringen85}.
Obviously, for $e_0=0$ we recover classical elasticity.
Eringen~\cite{Eringen83,Eringen85} used 
the value of $e_0=0.39$ in nonlocal elasticity.

We notice that a negative gradient coefficient $\kappa^{-2}<0$ would change
the character of the Helmholtz equations~(\ref{stress-fe}) and 
(\ref{strain-fe}) and the corresponding solutions. 
Let us emphasize that the solutions which we consider in the following
sections are valid for a positive gradient coefficient. 
In addition, non-negative definiteness of the strain energy~(\ref{en-sy}),
$W\geq 0$, requires the following conditions for the material constants and 
the gradient coefficient (see, e.g., \cite{GVV04})
\begin{align}
3\lambda+2\mu\geq 0,\qquad
\mu\geq 0,\qquad
\kappa^{-2}\geq 0.
\end{align}

\section{Dislocations}
In this section we consider straight dislocations whose line
coincides with the $z$-axis of a Cartesian coordinate system in an 
infinitely extended medium.

\subsection{Screw dislocation: $\burger=(0,0,b_z)$}
\begin{figure}[t]\unitlength1cm
\centerline{
\epsfig{figure=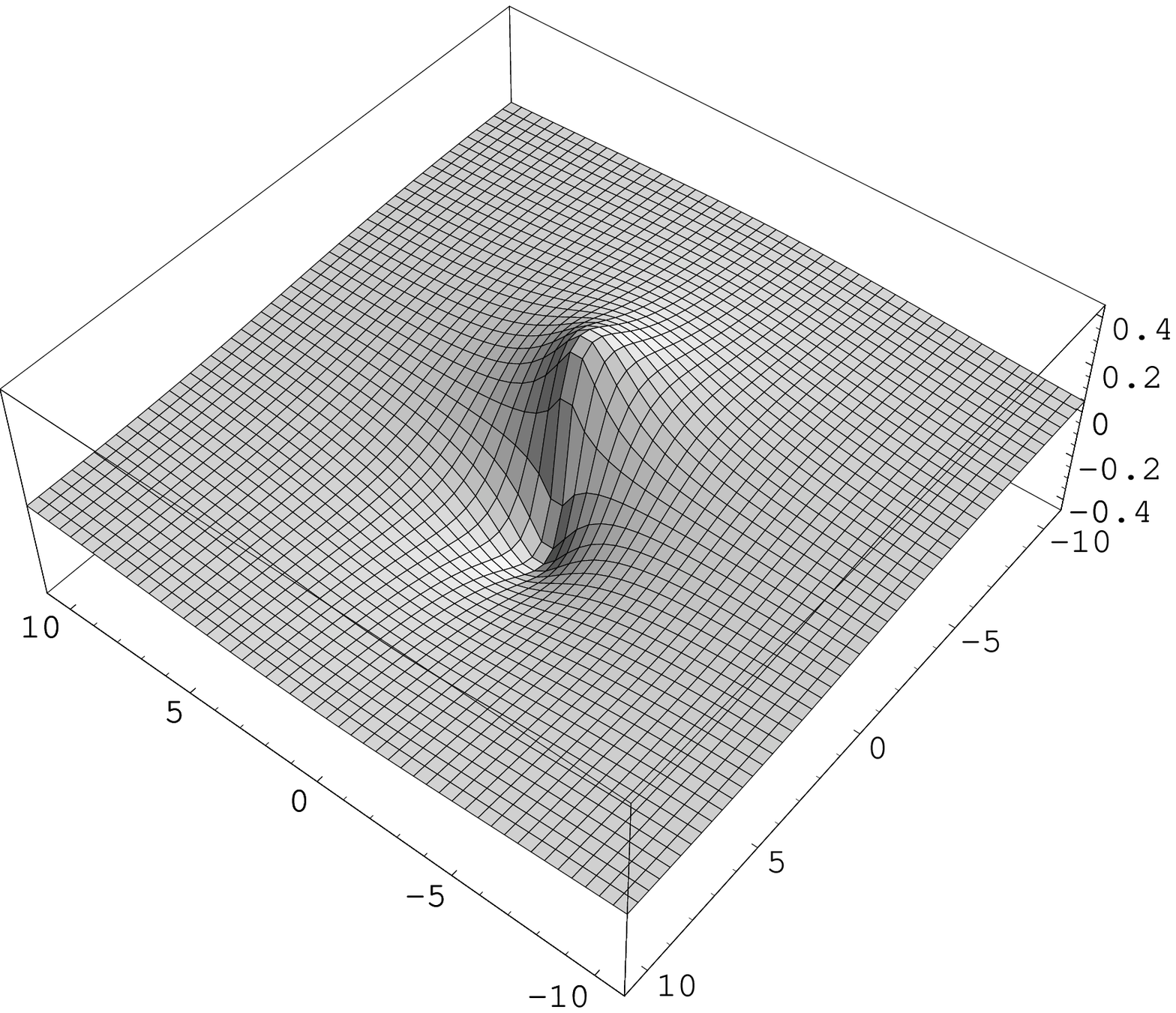,width=7.0cm}
\put(-1.0,1.0){$\kappa y$}
\put(-6.0,1.0){$\kappa x$}
\put(-6.2,-0.3){$\text{(a)}$}
\hspace*{0.4cm}
\put(0,-0.3){$\text{(b)}$}
\put(6.0,1.0){$\kappa y$}
\put(1.0,1.0){$\kappa x$}
\epsfig{figure=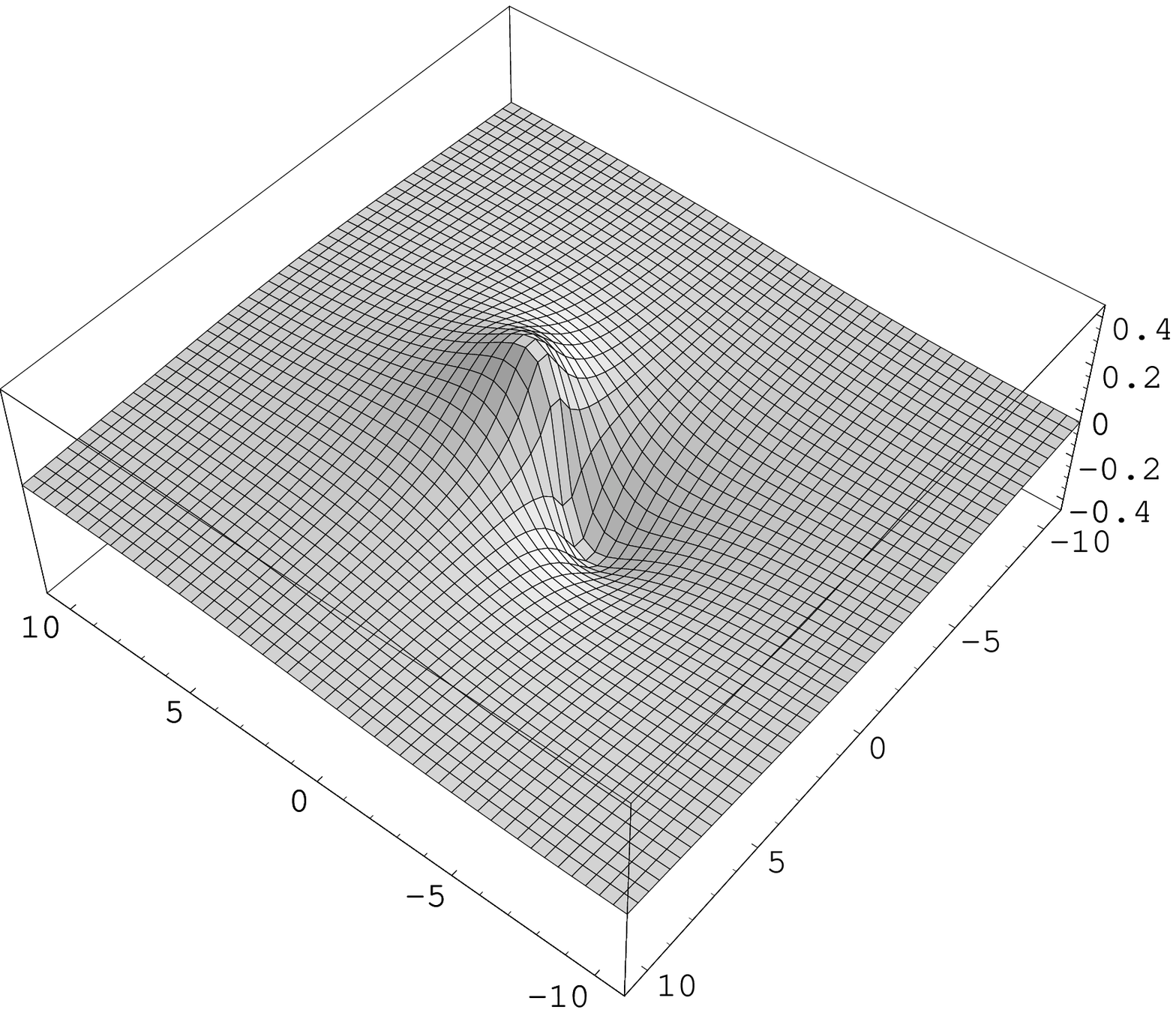,width=7.0cm}
}
\caption{Stress of a screw dislocation near the dislocation line:
(a) $\sigma_{xz}$ and  
(b) $\sigma_{yz}$
are given in units of $\mu b_z\kappa/[2\pi]$ .}
\label{fig:stress-screw}
\end{figure}

We start with the simplest case, the anti-plane strain which
corresponds to a screw dislocation. 
We make an ansatz for the total stress and for the Cauchy stress which 
has the form as
\begin{align}
\label{SFA-screw}
\tl\sigma {}_{ij}=
\left(\begin{array}{ccc}
0 & 0 & -\pd_{y} \tl F \\
0 & 0 & \pd_{x} \tl F\\
-\pd_{y} \tl F  &\pd_{x} \tl F  & 0
\end{array}\right),\quad
\sigma_{ij}=
\left(\begin{array}{ccc}
0 & 0 & -\pd_{y} F \\
0 & 0 & \pd_{x} F\\
-\pd_{y} F  &\pd_{x} F  & 0
\end{array}\right).
\end{align}
We choose for $\tl F$  the well-known stress function of elastic torsion, sometimes 
called Prandtl's stress function. 
It is given by (see, e.g.,~\cite{Kroener81})
\begin{align}
\label{SF-Pr}
\tl F=\frac{\mu b_z}{2\pi}\, \ln r,
\end{align}
with $r=\sqrt{x^2+y^2}$.
Substituting~(\ref{SFA-screw}) and (\ref{SF-Pr}) into~(\ref{stress-fe}) 
we obtain for the stress function $F$ the following 
inhomogeneous Helmholtz equation
\begin{align}
\label{HE-F}
\Big(1-\kappa^{-2}\Delta\Big)F=\frac{\mu b_z}{2\pi}\, \ln r.
\end{align}
The nonsingular solution of~(\ref{HE-F}) is (see, e.g.,~\cite{Eringen90,Edelen96,Lazar02b})
\begin{align}
\label{SF-F}
F=\frac{\mu b_z}{2\pi}\Big\{\ln r +K_0(\kappa r)\Big\},
\end{align}
which represents a stress function for a nonsingular screw dislocation. 
In the far field, the stress function~(\ref{SF-F}) agrees
with Prandtl's stress function and for small $r$ it cancels the logarithmic
singularity.  
Consequently, the elastic stress is given by 
\begin{align}
\label{T-screw}
\sigma_{zx}=-\frac{\mu b_z}{2\pi}\,\frac{y}{r^2}\Big\{1-\kappa r K_1(\kappa r)\Big\},
\qquad
\sigma_{zy}=\frac{\mu b_z}{2\pi}\,\frac{x}{r^2}\Big\{1-\kappa r K_1(\kappa r)\Big\},
\end{align}
the corresponding field of elastic strains reads
\begin{align}
\label{E-screw}
E_{zx}=-\frac{b_z}{4\pi}\,\frac{y}{r^2}\Big\{1-\kappa r K_1(\kappa r)\Big\},
\qquad
E_{zy}=\frac{b_z}{4\pi}\,\frac{x}{r^2}\Big\{1-\kappa r K_1(\kappa r)\Big\}.
\end{align}
The appearance of the modified Bessel function in~(\ref{T-screw}) and 
(\ref{E-screw})
leads to the elimination of classical singularity $\sim r^{-1}$ 
at the dislocation line (see Eq.~(\ref{K1-exp}) in Appendix~\ref{append}).
The stress $\sigma_{zy}$ has its extreme value 
$|\sigma_{zy}(x,0)|\simeq 0.399\kappa \frac{\mu b_z}{2\pi}$ at 
$|x|\simeq 1.114 \kappa^{-1}$,
whereas the stress $\sigma_{zx}$ has its extreme value 
$|\sigma_{zx}(0,y)|\simeq 0.399\kappa \frac{\mu b_z}{2\pi}$ at 
$|y|\simeq 1.114 \kappa^{-1}$.
The strain~$E_{zy}$ has its extreme value 
$|E_{zy}(x,0)|\simeq 0.399\kappa \frac{b_z}{4\pi}$ at 
$|x|\simeq 1.114 \kappa^{-1}$ and
 $E_{zx}$ has its extreme value 
$|E_{zx}(0,y)|\simeq 0.399\kappa \frac{b_z}{4\pi}$ at 
$|y|\simeq 1.114 \kappa^{-1}$. 
In addition, the stresses and strains are zero at $r=0$.  
The stress is plotted in
Fig.~\ref{fig:stress-screw}.
It is worth to note that the stress~(\ref{T-screw}) agrees with the 
nonlocal stress given by Eringen~\cite{Eringen83,Eringen85,Eringen90,Eringen02},
with Edelen's expression~\cite{Edelen96} and with the stress given by
Gutkin and Aifantis~\cite{GA99,GA00,Gutkin00}.  

Let us now give some remarks on the double stresses of the screw dislocation.
The non-vanishing components are given by (and the components due to the 
symmetry
$\tau_{(ij)k}$)
\begin{align}
\tau_{zyx}=\frac{1}{\kappa^2}\,\pd^2_{xx}F,\qquad
\tau_{zxy}=-\frac{1}{\kappa^2}\,\pd^2_{yy}F,\qquad
\tau_{zyy}=-\tau_{zxx}=\frac{1}{\kappa^2}\,\pd^2_{xy}F.
\end{align}
They have a similar form like the elastic bend-twist tensor of
a screw dislocation given in~\cite{Lazar02d}. 
This means that the double stresses are singular at $r=0$. 
Because the double stress is a simple gradient of the Cauchy stress
which is not singular in our case, it is less singular as a gradient of the
stress calculated in classical elasticity.

\subsection{Edge dislocation: $\burger=(b_x,0,0)$}
\begin{figure}[t]\unitlength1cm
\centerline{
\epsfig{figure=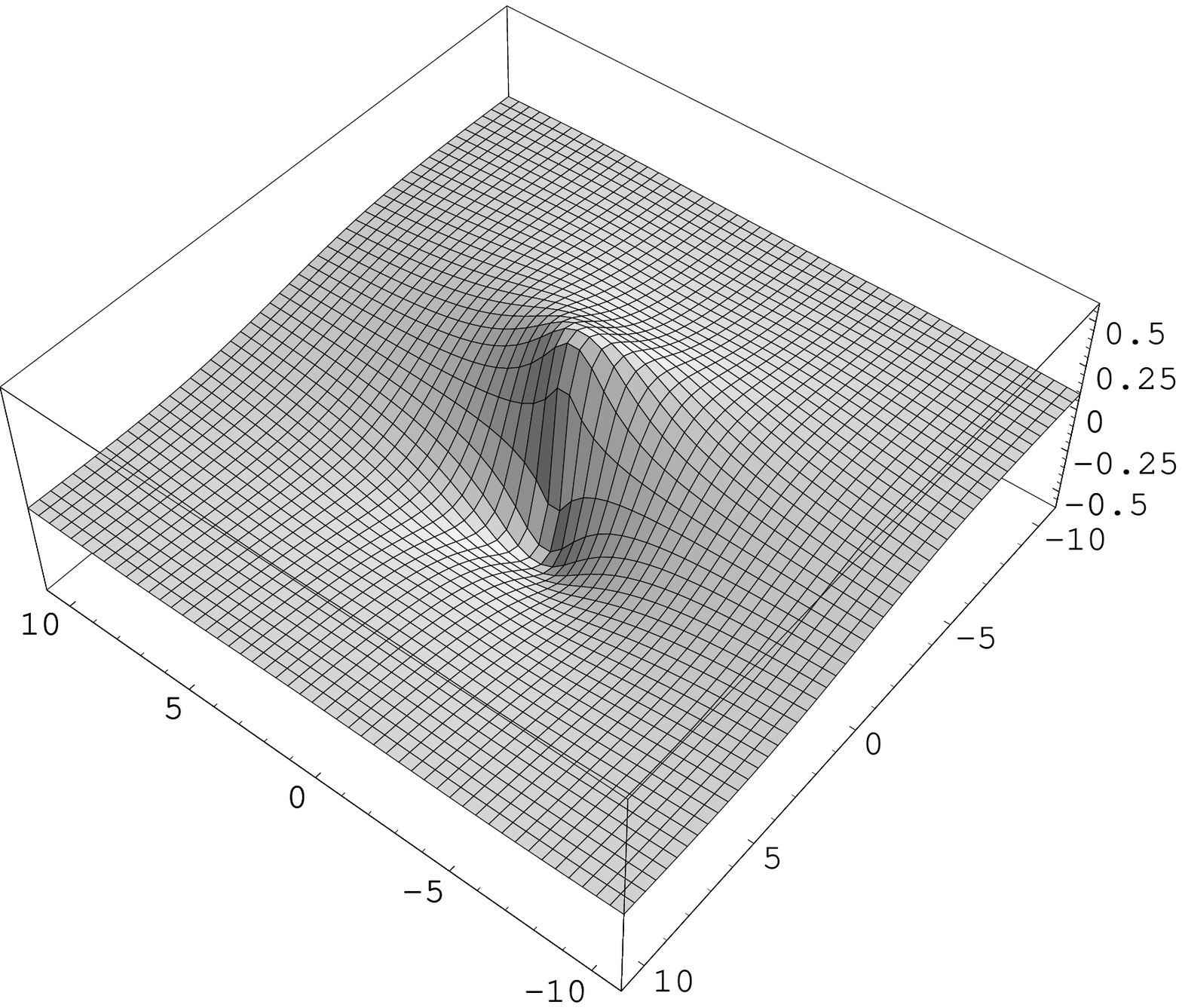,width=7.0cm}
\put(-1.0,1.0){$\kappa y$}
\put(-6.0,1.0){$\kappa x$}
\put(-6.2,-0.3){$\text{(a)}$}
\hspace*{0.4cm}
\put(0,-0.3){$\text{(b)}$}
\put(6.0,1.0){$\kappa y$}
\put(1.0,1.0){$\kappa x$}
\epsfig{figure=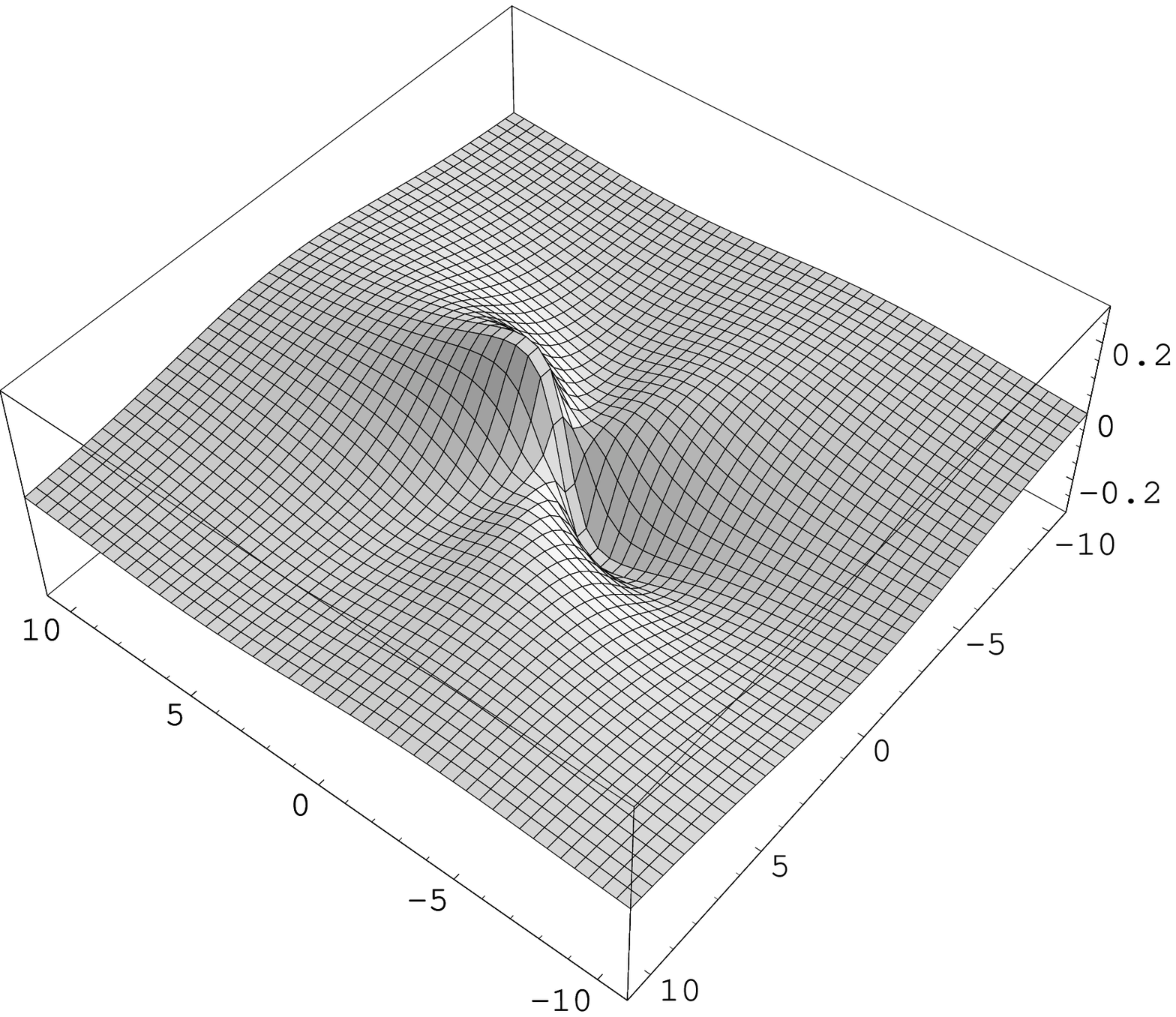,width=7.0cm}
}
\vspace*{0.2cm}
\centerline{
\epsfig{figure=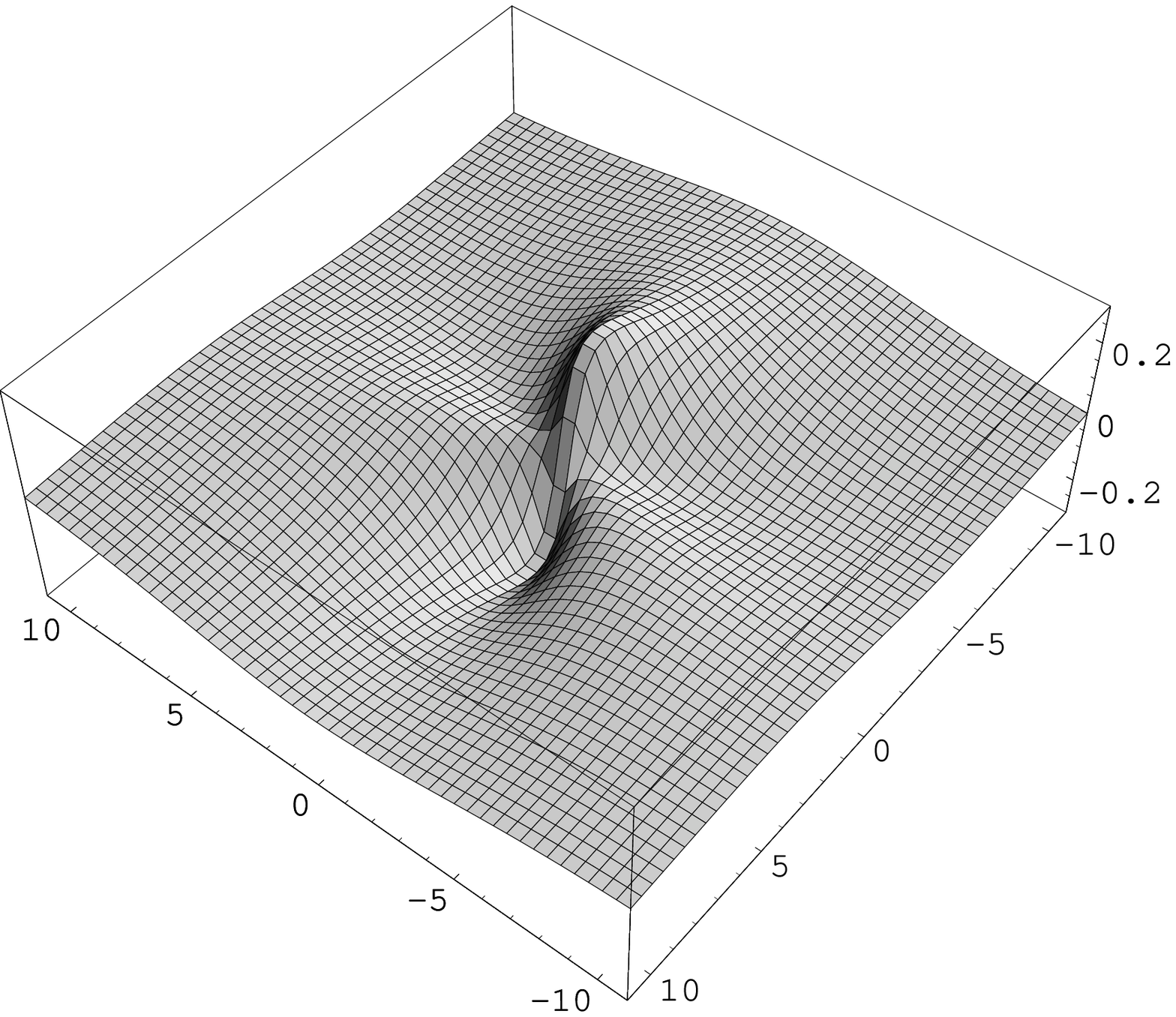,width=7.0cm}
\put(-1.0,1.0){$\kappa y$}
\put(-6.0,1.0){$\kappa x$}
\put(-6.0,-0.3){$\text{(c)}$}
\hspace*{0.4cm}
\put(6.0,1.0){$\kappa y$}
\put(1.0,1.0){$\kappa x$}
\put(0,-0.3){$\text{(d)}$}
\epsfig{figure=Tzz-x.eps,width=7.0cm}
}
\caption{Stress of an edge dislocation:
(a) $\sigma_{xx}$,
(b) $\sigma_{xy}$,
(c) $\sigma_{yy}$
are given in units of $\mu b_x\kappa/[2\pi(1-\nu)]$ 
and
(d) $\sigma_{zz}$
is given in units of $\mu b_x\nu \kappa/[\pi(1-\nu)]$ 
.}
\label{fig:stress-edge-x}
\end{figure}

In the case of plane strain, we may make the following stress function ansatz 
\begin{align}
\label{stress-ansatz}
\tl\sigma {}_{ij}=
\left(\begin{array}{ccc}
\pd^2_{yy}\tl f & -\pd^2_{xy}\tl f & 0\\
-\pd^2_{xy}\tl f & \pd^2_{xx}\tl f & 0\\
0& 0& \nu\Delta \tl f
\end{array}\right),\quad
\sigma_{ij}=
\left(\begin{array}{ccc}
\pd^2_{yy}f & -\pd^2_{xy}f & 0\\
-\pd^2_{xy}f & \pd^2_{xx}f & 0\\
0& 0& \nu\Delta f
\end{array}\right).
\end{align}
Obviously, it yields $\sigma_{zz}=\nu(\sigma_{xx}+\sigma_{yy})$.
For an edge dislocation with Burgers vector $b_x$, we use the corresponding
Airy's stress function~\cite{Kroener81}
\begin{align}
\label{Airy1}
\tl f=-\frac{\mu b_x}{2\pi(1-\nu)}\, y\ln r.
\end{align}
If we substitute (\ref{stress-ansatz}) and (\ref{Airy1})
into (\ref{stress-fe}), we get the inhomogeneous Helmholtz equation 
for the stress function $f$ 
\begin{align}
\label{f_fe_edge_x}
\Big(1-\kappa^{-2}\Delta\Big)f=- \frac{\mu b_x}{2\pi(1-\nu)}\, y\ln r .
\end{align}
The nonsingular solution for the modified stress function of a straight edge dislocation
is given by~\cite{Lazar03a,Lazar02d}
\begin{align}
\label{f-edge-x}
f=-\frac{\mu b_x}{2\pi(1-\nu)}\, y \bigg\{\ln r 
+\frac{2}{\kappa^2 r^2}\Big(1-\kappa r K_1(\kappa r)\Big)\bigg\}. 
\end{align}
By means of Eqs.~(\ref{stress-ansatz}) and (\ref{f-edge-x}),
the elastic stress is given as~\cite{Lazar03a}
\begin{align}
&\sigma_{xx}=-\frac{\mu b_x}{2\pi(1-\nu)}\, 
\frac{y}{r^4}\bigg\{\big(y^2+3x^2\big)+\frac{4}{\kappa^2r^2}\big(y^2-3x^2\big)
-2 y^2\kappa r K_1(\kappa r)-2\big(y^2-3x^2\big) K_2(\kappa r)\bigg\},\nonumber\\
&\sigma_{yy}=-\frac{\mu b_x}{2\pi(1-\nu)}\, 
\frac{y}{r^4}\bigg\{\big(y^2-x^2\big)-\frac{4}{\kappa^2r^2}\big(y^2-3x^2\big)
-2 x^2\kappa r K_1(\kappa r)+2\big(y^2-3x^2\big) K_2(\kappa r)\bigg\},\nonumber\\
&\sigma_{xy}=\frac{\mu b_x}{2\pi(1-\nu)}\, 
\frac{x}{r^4}\bigg\{\big(x^2-y^2\big)-\frac{4}{\kappa^2r^2}\big(x^2-3y^2\big)
-2 y^2\kappa r K_1(\kappa r)+2\big(x^2-3y^2\big) K_2(\kappa r)\bigg\},\nonumber\\
\label{T-edge-x}
&\sigma_{zz}=-\frac{\mu b_x \nu }{\pi(1-\nu)}\, 
\frac{y}{r^2}\Big\{1-\kappa r K_1(\kappa r)\Big\}.
\end{align}
The stress~(\ref{T-edge-x}) is zero at $r=0$.  
In fact, the ``classical'' singularities are eliminated due to the 
behaviour of the Bessel functions at the dislocation line (see Appendix~\ref{append}).
The stress~(\ref{T-edge-x}) has the
following extreme values:
$|\sigma_{xx}(0,y)|\simeq 0.546\kappa \frac{\mu b_x}{2\pi(1-\nu)}$ at 
$|y|\simeq 0.996 \kappa^{-1}$,
$|\sigma_{yy}(0,y)|\simeq 0.260 \kappa\frac{\mu b_x}{2\pi(1-\nu)}$ at 
$|y|\simeq 1.494 \kappa^{-1}$,
$|\sigma_{xy}(x,0)|\simeq 0.260 \kappa\frac{\mu b_x}{2\pi(1-\nu)}$ at 
$|x|\simeq 1.494 \kappa^{-1}$,
and
$|\sigma_{zz}(0,y)|\simeq 0.399\kappa \frac{\mu b_x\nu}{\pi(1-\nu)}$ at 
$|y|\simeq 1.114 \kappa^{-1}$.
The stress is plotted in Fig.~\ref{fig:stress-edge-x}.
The corresponding trace of the stress tensor $\sigma_{kk}=\sigma_{xx}+\sigma_{yy}+\sigma_{zz}$ 
produced by the edge dislocation is
\begin{align}
\label{}
\sigma_{kk}=-\frac{\mu b_x(1+\nu)}{\pi(1-\nu)}\, 
\frac{y}{r^2}\Big\{1-\kappa  r K_1(\kappa r)\Big\}.
\end{align}
Using the inverse of Hooke's law, we find
for the elastic strain of this edge dislocation 
\begin{align}
\label{E-edge-x}
&E_{xx}=-\frac{b_x}{4\pi(1-\nu)}\, 
\frac{y}{r^2}
\bigg\{(1-2\nu)+\frac{2x^2}{r^2}+\frac{4}{\kappa^2r^4}\big(y^2-3x^2\big)\nonumber\\
&\hspace{4.5cm}-2\left(\frac{y^2}{r^2}-\nu\right)\kappa r K_1(\kappa r)
-\frac{2}{r^2}\big(y^2-3x^2\big) K_2(\kappa r)\bigg\},\\
&E_{yy}=-\frac{b_x}{4\pi(1-\nu)}\, 
\frac{y}{r^2}
\bigg\{(1-2\nu)-\frac{2x^2}{r^2}-\frac{4}{\kappa^2r^4}\big(y^2-3x^2\big)\nonumber\\
&\hspace{4.5cm}-2\left(\frac{x^2}{r^2}-\nu\right)\kappa r K_1(\kappa r)
+\frac{2}{r^2}\big(y^2-3x^2\big) K_2(\kappa r)\bigg\},\nonumber\\
&E_{xy}=\frac{b_x}{4\pi(1-\nu)}\, 
\frac{x}{r^2}
\bigg\{1-\frac{2y^2}{r^2}-\frac{4}{\kappa^2r^4}\big(x^2-3y^2\big)
-\frac{2y^2}{r^2}\,\kappa r K_1(\kappa r)
+\frac{2}{r^2}\big(x^2-3y^2\big) K_2(\kappa r)\bigg\}.\nonumber
\end{align}
The strain~(\ref{E-edge-x}) has the
extreme values ($\nu=0.3$):
$|E_{xx}(0,y)|\simeq 0.308\kappa \frac{ b_x}{4\pi(1-\nu)}$ at 
$|y|\simeq 0.922 \kappa^{-1}$,
$|E_{yy}(0,y)|\simeq 0.010 \kappa\frac{ b_x}{4\pi(1-\nu)}$ at 
$|y|\simeq 0.218 \kappa^{-1}$, 
$|E_{yy}(0,y)|\simeq 0.054 \kappa\frac{ b_x}{4\pi(1-\nu)}$ at 
$|y|\simeq 4.130 \kappa^{-1}$, 
and
$|E_{xy}(x,0)|\simeq 0.260 \kappa\frac{ b_x}{4\pi(1-\nu)}$ at 
$|x|\simeq 1.494 \kappa^{-1}$. 
In addition, it is interesting to note that 
$E_{yy}(0,y)$ is much smaller than $E_{xx}(0,y)$ within
the core region (see also~\cite{GA99}).
The dilatation $E_{kk}$ reads
\begin{align}
E_{kk}=-\frac{b_x(1-2\nu)}{2\pi(1-\nu)}\, 
\frac{y}{r^2}\Big\{1-\kappa  r K_1(\kappa r)\Big\}.
\end{align}
The strain and the dilatation are zero at the dislocation line.

It is interesting to note that   
the stress~(\ref{T-edge-x}) and the strain~(\ref{E-edge-x}) of a 
dislocation with Burgers vector $b_x$ agree with the 
expressions calculated by Gutkin and Aifantis~\cite{GA99,GA00,Gutkin00} 
by using the technique of Fourier transformation.

\subsection{Edge dislocation: $\burger=(0,b_y,0)$}
Let us now complete the case of dislocations with the edge dislocation with 
Burgers vector $b_y$. 
Again it is a plane strain state such that we can use the stress function 
ansatz~(\ref{stress-ansatz}).
But now Airy's stress function reads 
\begin{align}
\label{Airy2}
\tl f=\frac{\mu b_y}{2\pi(1-\nu)}\, x\ln r.
\end{align}
Substituting (\ref{stress-ansatz}) and (\ref{Airy2})
into (\ref{stress-fe}) we find 
\begin{align}
\label{f_fe_edge_y}
\Big(1-\kappa^{-2}\Delta\Big)f= \frac{\mu b_y}{2\pi(1-\nu)}\, x\ln r .
\end{align}
Its solution is given by
\begin{align}
\label{f-edge-y}
f=\frac{\mu b_y}{2\pi(1-\nu)}\, x \bigg\{\ln r 
+\frac{2}{\kappa^2 r^2}\Big(1-\kappa r K_1(\kappa r)\Big)\bigg\}.
\end{align}
Then we find for the elastic stress
\begin{align}
&\sigma_{xx}=\frac{\mu b_y}{2\pi(1-\nu)}\, 
\frac{x}{r^4}\bigg\{\big(x^2-y^2\big)-\frac{4}{\kappa^2r^2}\big(x^2-3y^2\big)
-2 y^2\kappa r K_1(\kappa r)+2\big(x^2-3y^2\big) K_2(\kappa r)\bigg\},\nonumber\\
&\sigma_{yy}=\frac{\mu b_y}{2\pi(1-\nu)}\, 
\frac{x}{r^4}\bigg\{\big(x^2+3y^2\big)+\frac{4}{\kappa^2r^2}\big(x^2-3y^2\big)
-2 x^2\kappa r K_1(\kappa r)-2\big(x^2-3y^2\big) K_2(\kappa r)\bigg\},\nonumber\\
&\sigma_{xy}=\frac{\mu b_y}{2\pi(1-\nu)}\, 
\frac{y}{r^4}\bigg\{\big(x^2-y^2\big)-\frac{4}{\kappa^2r^2}\big(3x^2-y^2\big)
+2 x^2\kappa r K_1(\kappa r)+2\big(3x^2-y^2\big) K_2(\kappa r)\bigg\},\nonumber\\
\label{T-edge-y}
&\sigma_{zz}=\frac{\mu b_y \nu }{\pi(1-\nu)}\, 
\frac{x}{r^2}\Big\{1-\kappa r K_1(\kappa r)\Big\},
\end{align}
and for the trace of the stress tensor
\begin{align}
\label{}
\sigma_{kk}=\frac{\mu b_y(1+\nu)}{\pi(1-\nu)}\, 
\frac{x}{r^2}\Big\{1-\kappa  r K_1(\kappa r)\Big\}.
\end{align}
The stress~(\ref{T-edge-y}) has 
following extreme values:
$|\sigma_{yy}(x,0)|\simeq 0.546\kappa \frac{\mu b_y}{2\pi(1-\nu)}$ at 
$|x|\simeq 0.996 \kappa^{-1}$,
$|\sigma_{xx}(x,0)|\simeq 0.260 \kappa\frac{\mu b_y}{2\pi(1-\nu)}$ at 
$|x|\simeq 1.494 \kappa^{-1}$,
$|\sigma_{xy}(0,y)|\simeq 0.260 \kappa\frac{\mu b_y}{2\pi(1-\nu)}$ at 
$|y|\simeq 1.494 \kappa^{-1}$,
and
$|\sigma_{zz}(x,0)|\simeq 0.399\kappa \frac{\mu b_y\nu}{\pi(1-\nu)}$ at 
$|x|\simeq 1.114 \kappa^{-1}$.

The elastic strain is given by
\begin{align}
\label{E-edge-y}
&E_{xx}=\frac{b_y}{4\pi(1-\nu)}\, 
\frac{x}{r^2}
\bigg\{(1-2\nu)-\frac{2y^2}{r^2}-\frac{4}{\kappa^2r^4}\big(x^2-3y^2\big)\nonumber\\
&\hspace{4.5cm}-2\left(\frac{y^2}{r^2}-\nu\right)\kappa r K_1(\kappa r)
+\frac{2}{r^2}\big(x^2-3y^2\big) K_2(\kappa r)\bigg\},\\
&E_{yy}=\frac{b_y}{4\pi(1-\nu)}\, 
\frac{x}{r^2}
\bigg\{(1-2\nu)+\frac{2y^2}{r^2}+\frac{4}{\kappa^2r^4}\big(x^2-3y^2\big)\nonumber\\
&\hspace{4.5cm}-2\left(\frac{x^2}{r^2}-\nu\right)\kappa r K_1(\kappa r)
-\frac{2}{r^2}\big(x^2-3y^2\big) K_2(\kappa r)\bigg\},\nonumber\\
&E_{xy}=-\frac{b_y}{4\pi(1-\nu)}\, 
\frac{y}{r^2}
\bigg\{1-\frac{2x^2}{r^2}+\frac{4}{\kappa^2r^4}\big(3x^2-y^2\big)
-\frac{2x^2}{r^2}\,\kappa r K_1(\kappa r)
-\frac{2}{r^2}\big(3x^2-y^2\big) K_2(\kappa r)\bigg\},\nonumber
\end{align}
and the dilatation reads
\begin{align}
E_{kk}=\frac{b_y(1-2\nu)}{2\pi(1-\nu)}\, 
\frac{x}{r^2}\Big\{1-\kappa  r K_1(\kappa r)\Big\}.
\end{align}
The strain~(\ref{E-edge-y}) has the
extreme values ($\nu=0.3$):
$|E_{yy}(x,0)|\simeq 0.308\kappa \frac{ b_y}{4\pi(1-\nu)}$ at 
$|x|\simeq 0.922 \kappa^{-1}$,
$|E_{xx}(x,0)|\simeq 0.010 \kappa\frac{ b_y}{4\pi(1-\nu)}$ at 
$|x|\simeq 0.218 \kappa^{-1}$, 
$|E_{xx}(x,0)|\simeq 0.054 \kappa\frac{ b_y}{4\pi(1-\nu)}$ at 
$|x|\simeq 4.130 \kappa^{-1}$, 
and
$|E_{xy}(0,y)|\simeq 0.260 \kappa\frac{ b_y}{4\pi(1-\nu)}$ at 
$|y|\simeq 1.494 \kappa^{-1}$. 
Again, it is interesting to note that 
$E_{xx}(x,0)$ is much smaller than $E_{yy}(x,0)$ within
the core region.
In addition, the stress and strain fields  
are zero at $r=0$.

In principle, we could calculate the double stresses of edge dislocations.
But we do not want to do this in detail.
Using Eqs.~(\ref{T-edge-x}) and (\ref{T-edge-y}) we would obtain expressions which are similar in form to
the double stresses of the screw dislocation. Again, the double stresses 
are still singular at the dislocation line.
The double stresses~\footnote{In the meantime~\cite{LMA04}, we have calculated 
the double and triple stresses of screw and edge dislocations in
second strain gradient elasticity. The double stresses can be found there
in the limit from second to first strain gradient elasticity.} 
of an edge dislocation are given in terms of the 
stress function $f$ 
as derivatives of the third order according to:
\begin{align}
\label{ds-edge1}
&\tau_{yyx}=\frac{1}{\kappa^2}\, \pd^3_{xxx} f ,\qquad\qquad
\tau_{yyy}=-\tau_{xyx}=\frac{1}{\kappa^2} \, \pd^3_{xxy} f \nonumber\\
&\tau_{xxy}=\frac{1}{\kappa^2} \, \pd^3_{yyy} f ,\qquad\qquad
\tau_{xxx}=-\tau_{xyy}=\frac{1}{\kappa^2} \, \pd^3_{yyx} f \nonumber\\
&\tau_{zzx}=\nu(\tau_{xxx}+\tau_{yyx})\qquad
\tau_{zzy}=\nu(\tau_{xxy}+\tau_{yyy})\, .
\end{align}
They are singular at $r=0$.

\section{Disclinations}
In this section we consider straight disclinations in an infinitely extended
medium. The disclination line
coincides with the $z$-axis of a Cartesian coordinate system.
We are using deWit's expressions~\cite{deWit73b} for the classical stress 
and strain fields (see also~\cite{RV}).

\subsection{Wedge disclination: $\frank=(0,0,\Omega_z)$}
\begin{figure}[t]\unitlength1cm
\centerline{
\epsfig{figure=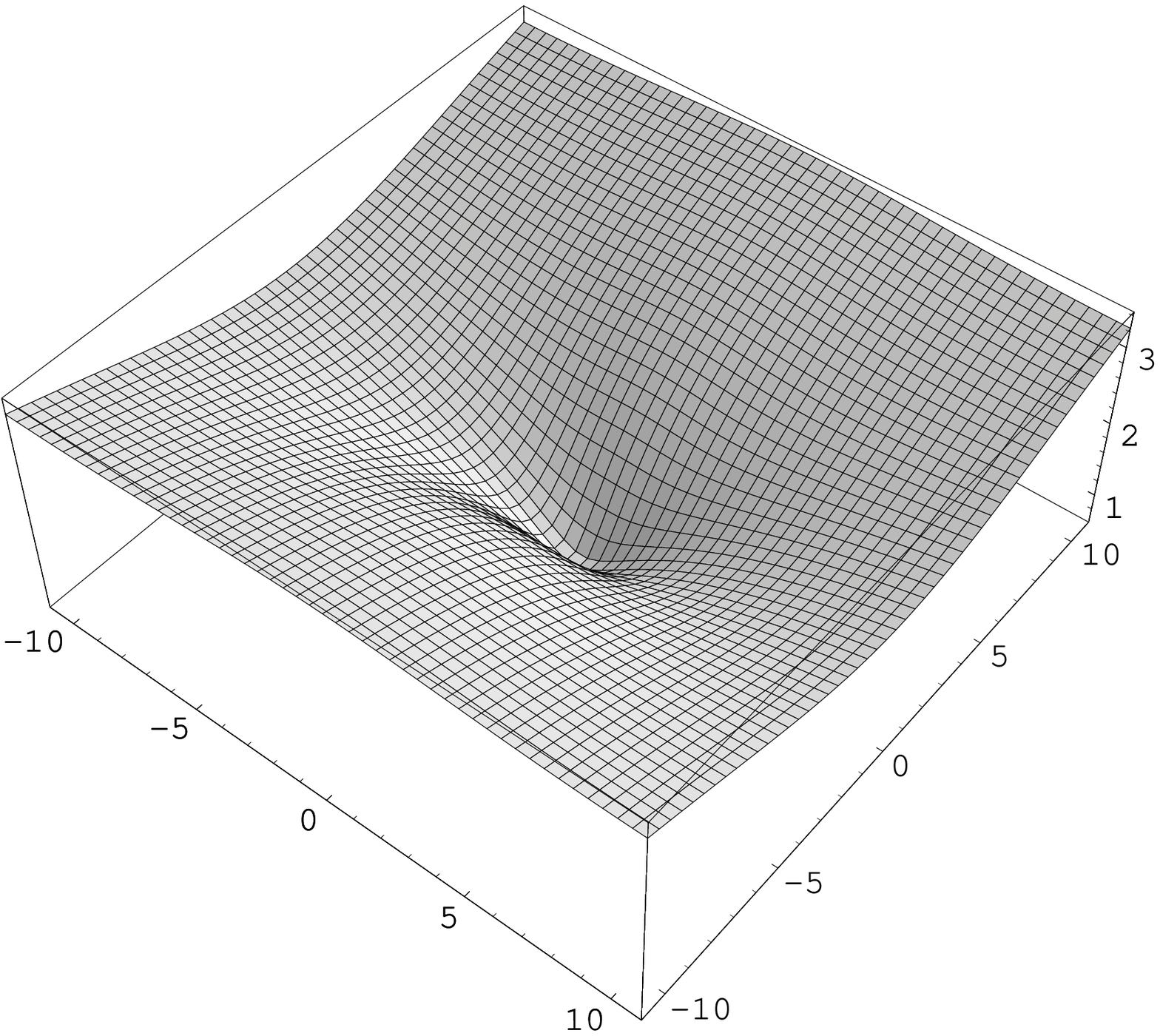,width=7.0cm}
\put(-1.0,1.0){$\kappa y$}
\put(-6.0,1.0){$\kappa x$}
\put(-6.2,-0.3){$\text{(a)}$}
\hspace*{0.4cm}
\put(0,-0.3){$\text{(b)}$}
\put(6.0,1.0){$\kappa y$}
\put(1.0,1.0){$\kappa x$}
\epsfig{figure=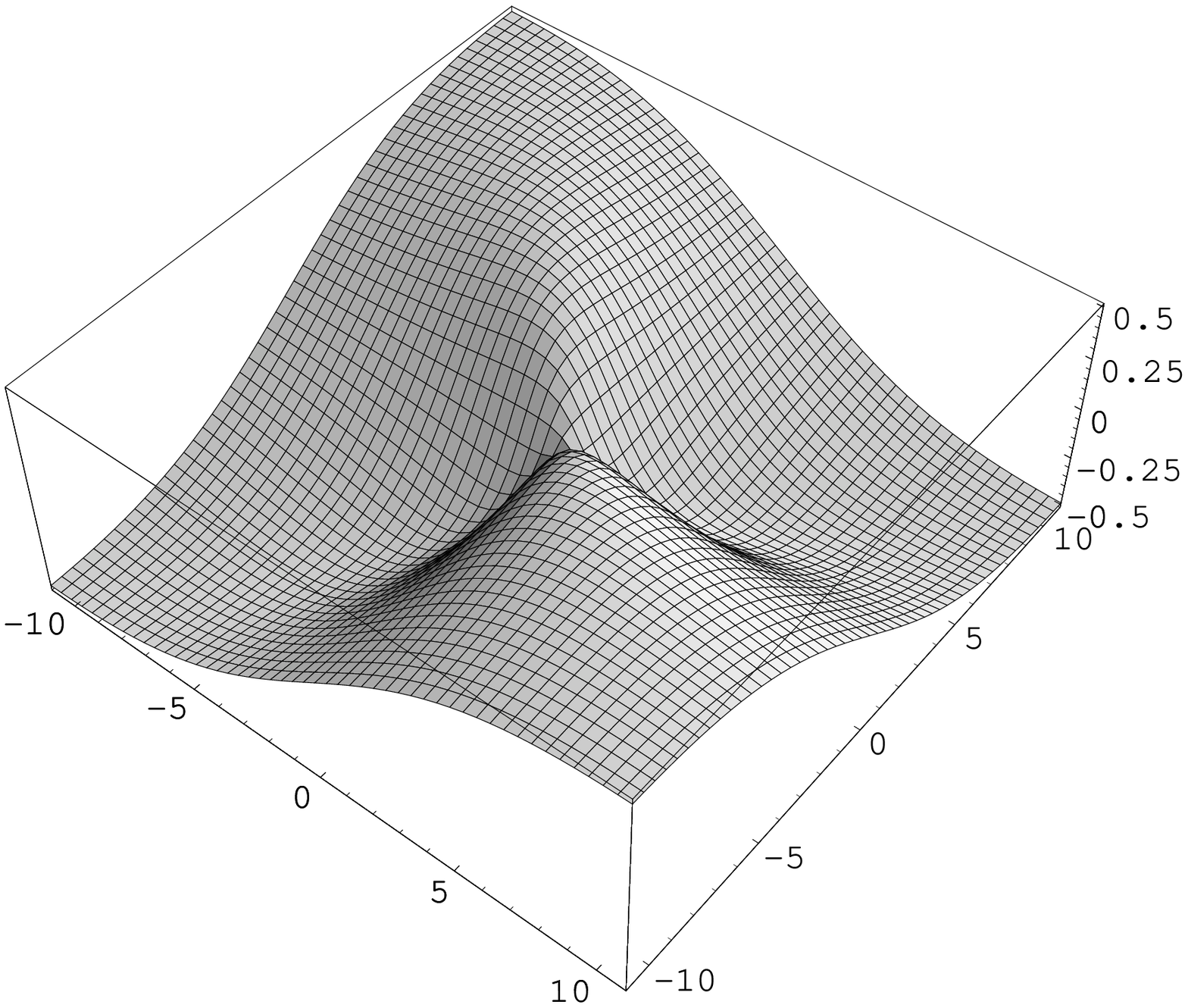,width=7.0cm}
}
\vspace*{0.2cm}
\centerline{
\epsfig{figure=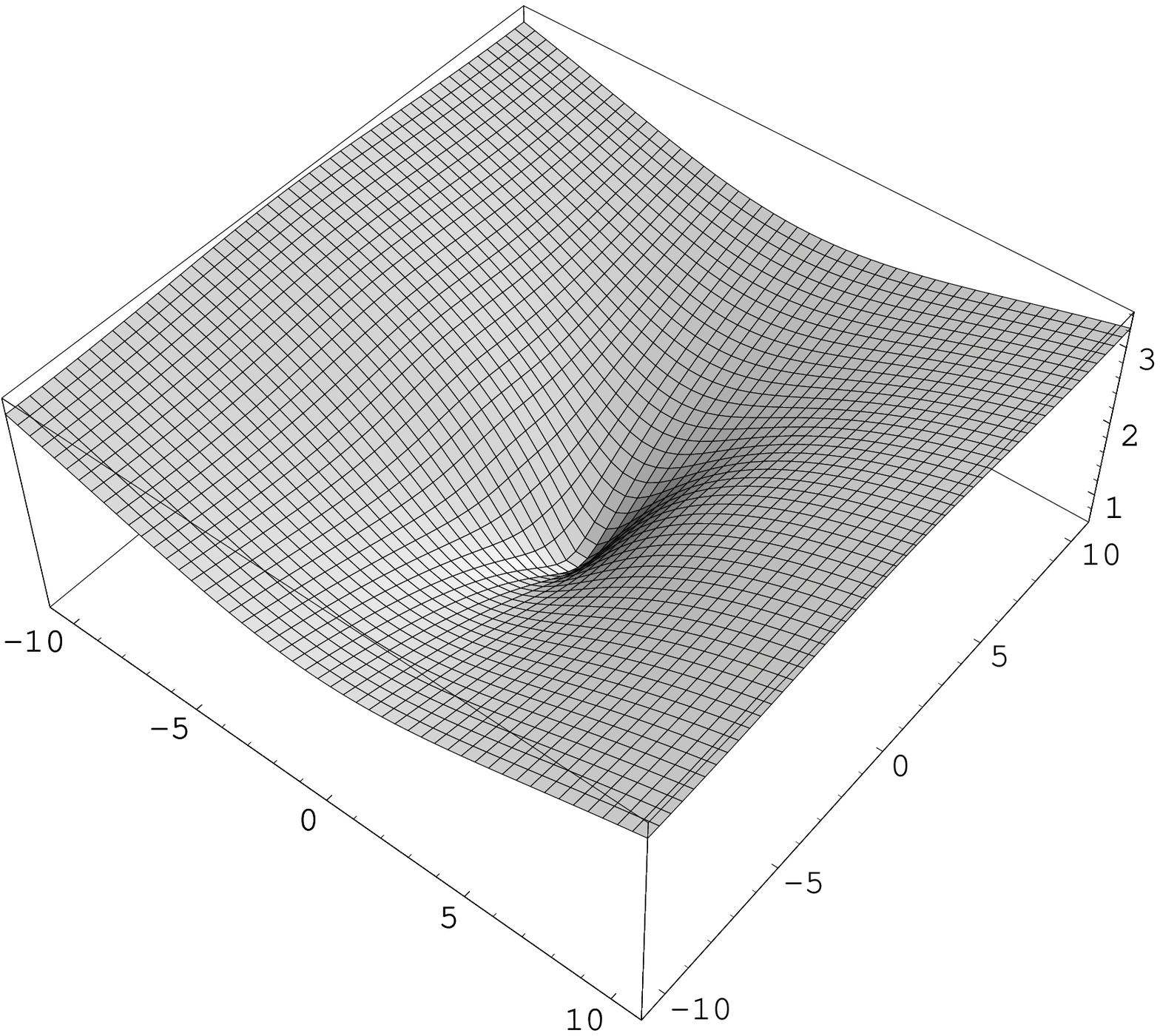,width=7.0cm}
\put(-1.0,1.0){$\kappa y$}
\put(-6.0,1.0){$\kappa x$}
\put(-6.0,-0.3){$\text{(c)}$}
\hspace*{0.4cm}
\put(6.0,1.0){$\kappa y$}
\put(1.0,1.0){$\kappa x$}
\put(0,-0.3){$\text{(d)}$}
\epsfig{figure=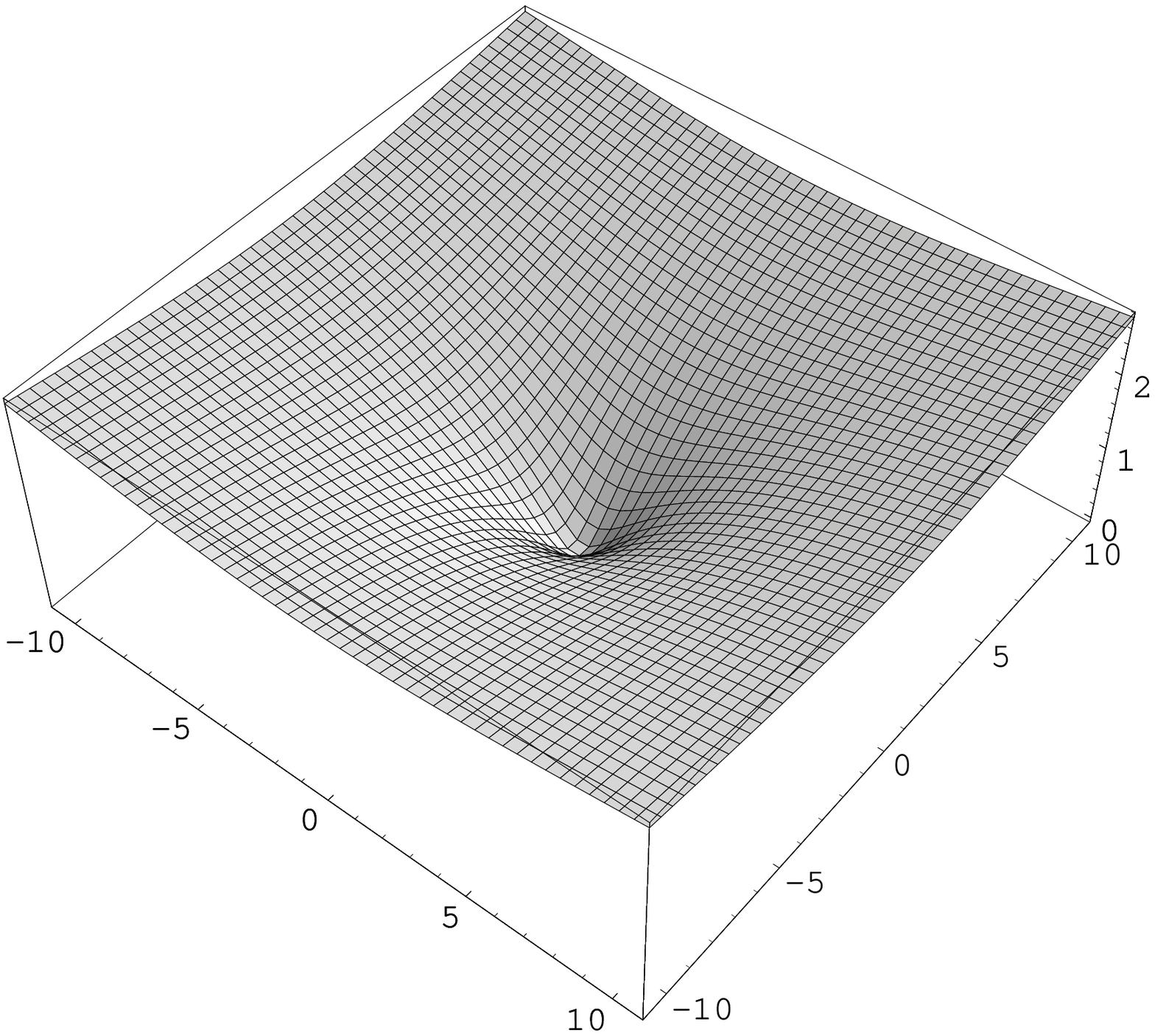,width=7.0cm}
}
\caption{Stress of a wedge disclination:
(a) $\sigma_{xx}$,
(b) $\sigma_{xy}$,
(c) $\sigma_{yy}$
are given in units of $\mu\Omega_z/[2\pi(1-\nu)]$ 
and
(d) $\sigma_{zz}$
is given in units of $\mu \Omega_z\nu/[\pi(1-\nu)]$ 
.}
\label{fig:stress-wedge}
\end{figure}

As in the case of edge dislocations we may use the stress function 
ansatz~(\ref{stress-ansatz}) for the wedge disclination as well. 
It is obvious because the wedge disclination corresponds also to 
plane strain. 
Using the stress function of a ``classical'' wedge disclination, 
\begin{align}
\label{SF-cl}
\tl f=\frac{\mu \Omega_z}{4\pi(1-\nu)}\, r^2\left\{ \ln r -\frac{1-4\nu}{2(1-2\nu)}\right\},
\end{align}
and (\ref{stress-ansatz}), we obtain the following equation 
which gives the solution of a ``modified'' stress function of a wedge
disclination 
\begin{align}
\label{f_fe_we}
\Big(1-\kappa^{-2}\Delta\Big)f=\frac{\mu \Omega_z}{4\pi(1-\nu)}\, r^2\left(\ln r -\frac{1-4\nu}{2(1-2\nu)}\right).
\end{align}
Consequently, the ``modified'' stress function of a wedge
disclination is given by (see also~\cite{Lazar03b})
\begin{align}
\label{f_wedge}
f=\frac{\mu \Omega_z}{4\pi(1-\nu)}\left\{r^2\bigg(\ln r-\frac{1-4\nu}{2(1-2\nu)}\bigg)
+\frac{4}{\kappa^2}\,\bigg(\ln r+K_0(\kappa r)+\frac{1}{2(1-2\nu)}\bigg)\right\}.
\end{align}
Substituting~(\ref{f_wedge}) into~(\ref{stress-ansatz}), we find for
the stresses of a wedge disclination
\begin{align}
\label{T-wedge}
&\sigma_{xx}=\frac{\mu \Omega_z}{2\pi(1-\nu)}\, 
\bigg\{\ln r+\frac{y^2}{r^2}+\frac{\nu}{1-2\nu} +K_0(\kappa r)
+\frac{\big(x^2-y^2\big)}{\kappa^2 r^4} \Big(2-\kappa^2 r^2 K_2(\kappa r)\Big)\bigg\},\nonumber\\
&\sigma_{yy}=\frac{\mu \Omega_z}{2\pi(1-\nu)}\, 
\bigg\{\ln r+\frac{x^2}{r^2}+\frac{\nu}{1-2\nu} +K_0(\kappa r)
-\frac{\big(x^2-y^2\big)}{\kappa^2 r^4} \Big(2-\kappa^2 r^2 K_2(\kappa r)\Big)\bigg\},\nonumber\\
&\sigma_{xy}=-\frac{\mu \Omega_z}{2\pi(1-\nu)}\, 
\frac{xy}{r^2}\bigg\{1-\frac{2}{\kappa^2 r^2} \Big(2-\kappa^2 r^2 K_2(\kappa r)\Big)\bigg\},\nonumber\\
&\sigma_{zz}=\frac{\mu \Omega_z\, \nu}{\pi(1-\nu)}\, 
\bigg\{\ln r+\frac{1}{2(1-2\nu)}+K_0(\kappa r)\bigg\},
\end{align}
and for the trace of the stress tensor
\begin{align}
\label{}
\sigma_{kk}=\frac{\mu \Omega_z(1+\nu)}{\pi(1-\nu)}\, 
\bigg\{\ln r+\frac{1}{2(1-2\nu)}+K_0(\kappa r)\bigg\}.
\end{align}
The stress is plotted in Fig.~\ref{fig:stress-wedge}.
Using Eqs.~(\ref{K0-exp}) and (\ref{K2-exp}), we obtain for the stress
at $r=0$
\begin{align}
&\sigma_{xx}(0)=\sigma_{yy}(0)=
\frac{\mu \Omega_z}{2\pi(1-\nu)}\, 
\bigg\{\frac{\nu}{1-2\nu}-\gamma-\ln \frac{\kappa }{2}+\frac{1}{2}\bigg\},
\quad\sigma_{xy}(0)=0,\nonumber\\
&\sigma_{zz}(0)=\frac{\nu}{1+\nu}\sigma_{kk}(0)=
\frac{\mu \Omega_z\nu}{\pi(1-\nu)}\, 
\bigg\{\frac{\nu}{1-2\nu}-\gamma-\ln \frac{\kappa }{2}\bigg\}.\nonumber
\end{align}
Consequently, the stress is finite at the disclination line in contrast to 
the unphysical stress singularity
in ``classical'' disclination theory. 
The elastic strain is easily calculated as
\begin{align}
\label{E-wedge}
&E_{xx}=\frac{\Omega_z}{4\pi(1-\nu)}
\bigg\{ (1-2\nu)\big(\ln r +K_0(\kappa r)\big)
+\frac{y^2}{r^2}+\frac{(x^2-y^2)}{\kappa^2 r^4}\Big(2-\kappa^2 r^2 K_2(\kappa r)\Big)\bigg\},\nonumber\\
&E_{yy}=\frac{\Omega_z}{4\pi(1-\nu)}
\bigg\{ (1-2\nu)\big(\ln r +K_0(\kappa r)\big)
+\frac{x^2}{r^2}-\frac{(x^2-y^2)}{\kappa^2 r^4}\Big(2-\kappa^2 r^2 K_2(\kappa r)\Big)\bigg\},\nonumber\\
&E_{xy}=-\frac{\Omega_z}{4\pi(1-\nu)}\, 
\frac{xy}{r^2}\bigg\{1-\frac{2}{\kappa^2 r^2} \Big(2-\kappa^2 r^2 K_2(\kappa r)\Big)\bigg\},
\end{align}
and for the dilatation we obtain
\begin{align}
\label{dilat-wedge}
E_{kk}=\frac{\Omega_z}{2\pi(1-\nu)}
\bigg\{(1-2\nu)\big(\ln r+K_0(\kappa r)\big)+\frac{1}{2}\bigg\}.
\end{align}
Again using Eqs.~(\ref{K0-exp}) and (\ref{K2-exp})
the strain reads at $r=0$
\begin{align}
E_{xx}(0)=E_{yy}(0)=\frac{1}{2} E_{kk}(0)=
-\frac{\Omega_z}{4\pi(1-\nu)}\, 
\bigg\{(1-2\nu)\Big(\gamma+\ln \frac{\kappa }{2}\Big)-\frac{1}{2}\bigg\},
\quad E_{xy}(0)=0.\nonumber
\end{align}

Now we calculate the non-vanishing components of the double stress for the
wedge disclination. They are given in terms of the stress function as follows 
\begin{align}
&\tau_{yyx}=\frac{1}{\kappa^2}\, \pd^3_{xxx}f,\qquad
\tau_{xxy}=\frac{1}{\kappa^2}\, \pd^3_{yyy}f,\qquad
\tau_{yyy}=-\tau_{xyx}=\frac{1}{\kappa^2}\, \pd^3_{yxx}f,\nonumber\\
&\tau_{xxx}=-\tau_{xyy}=\frac{1}{\kappa^2}\, \pd^3_{xyy}f,\qquad
\tau_{zzx}=\frac{\nu}{\kappa^2}\,\pd_x \Delta f,\qquad
\tau_{zzy}=\frac{\nu}{\kappa^2}\, \pd_y \Delta f.
\end{align}
So we obtain for the double stresses of the wedge disclination
\begin{align}
&\tau_{yyx}=\frac{\mu \Omega_z}{2\pi(1-\nu)\kappa^2}\, 
\frac{x}{r^4}\bigg\{\big(x^2+3y^2\big)+\frac{4}{\kappa^2r^2}\big(x^2-3y^2\big)
-2 x^2\kappa r K_1(\kappa r)-2\big(x^2-3y^2\big) K_2(\kappa r)\bigg\},\nonumber\\
&\tau_{xxy}=\frac{\mu \Omega_z}{2\pi(1-\nu)\kappa^2}\, 
\frac{y}{r^4}\bigg\{\big(y^2+3x^2\big)+\frac{4}{\kappa^2r^2}\big(y^2-3x^2\big)
-2 y^2\kappa r K_1(\kappa r)-2\big(y^2-3x^2\big) K_2(\kappa r)\bigg\},\nonumber\\
&\tau_{yyy}=\frac{\mu \Omega_z}{2\pi(1-\nu)\kappa^2}\, 
\frac{y}{r^4}\bigg\{\big(y^2-x^2\big)-\frac{4}{\kappa^2r^2}\big(y^2-3x^2\big)
-2 x^2\kappa r K_1(\kappa r)+2\big(y^2-3x^2\big) K_2(\kappa r)\bigg\},\nonumber\\
&\tau_{xxx}=\frac{\mu \Omega_z}{2\pi(1-\nu)\kappa^2}\, 
\frac{x}{r^4}\bigg\{\big(x^2-y^2\big)-\frac{4}{\kappa^2r^2}\big(x^2-3y^2\big)
-2 y^2\kappa r K_1(\kappa r)+2\big(x^2-3y^2\big) K_2(\kappa r)\bigg\},\nonumber\\
&\tau_{zzx}=\frac{\mu \Omega_z \nu }{\pi(1-\nu)\kappa^2}\, 
\frac{x}{r^2}\Big\{1-\kappa r K_1(\kappa r)\Big\},\nonumber\\
\label{DS-wedge}
&\tau_{zzy}=\frac{\mu \Omega_z \nu }{\pi(1-\nu)\kappa^2}\, 
\frac{y}{r^2}\Big\{1-\kappa r K_1(\kappa r)\Big\}.
\end{align}
This double stress tensor has some interesting features. 
First, it is not singular. Second, it has a similar form like the
stress field of an edge dislocation (compare with Eqs.~(\ref{T-edge-x}) and 
(\ref{T-edge-y})). Thus, the components are zero at $r=0$ and 
have extremum values near the disclination line.

\subsection{Twist disclination: $\frank=(\Omega_x,0,0)$}
In the case of twist disclination the problem, which we want to consider,
is more complicated than those of dislocations and of 
the wedge disclination. The 
reason is that the situation is no longer a proper two-dimensional problem.
In the case of twist disclination the three-dimensional space
may be considered as a product of the two-dimensional $xy$-plane and the 
independent one-dimensional $z$-line~\cite{deWit73b}. 
Thus, the $z$-axis plays a peculiar role.

First, we make an ansatz which fulfills the stress equilibrium. It is given
by 
\begin{align}
\label{stress-ansatz-tw1}
\tl\sigma {}_{ij}&=
\left(\begin{array}{ccc}
\ \,  \pd^2_{yy}\tl f & -\pd^2_{xy}\tl f & -\pd_y \tl F+\pd_z \tl g\\
-\pd^2_{xy}\tl f & \pd^2_{xx}\tl f & \pd_x \tl F\\
-\pd_y \tl F+\pd_z \tl g& \pd_x\tl F &\tl p
\end{array}\right),\ \nonumber\\
\sigma_{ij}&=
\left(\begin{array}{ccc}
\ \,  \pd^2_{yy}f & -\pd^2_{xy}f & -\pd_y F+\pd_z g\\
-\pd^2_{xy}f & \pd^2_{xx}f & \pd_x F \\
-\pd_y F+\pd_z g& \pd_x F& p
\end{array}\right),
\end{align}
with the relations
\begin{align}
\tl p=\nu\Delta \tl f=-\pd_x \tl g,\qquad
p=\nu\Delta f=-\pd_x g,
\end{align}
which follow from $\pd_i\tl\sigma_{zi}=0$ and $\pd_i\sigma_{zi}=0$.
One can see the special role of $z$ in the ansatz~(\ref{stress-ansatz-tw1}).
The ansatz~(\ref{stress-ansatz-tw1}) is not only an addition of the
anti-plane~(\ref{SFA-screw}) and plane strain~(\ref{stress-ansatz}) situation 
because an additional stress function $\tl g$ or $g$ enters the ansatz. 
The following ``classical'' stress functions,
\begin{align}
\label{f_twist_cl1}
&\tl f=-\frac{\mu \Omega_x}{2\pi(1-\nu)}\, z x \ln r,\nonumber\\
&\tl F=\frac{\mu \Omega_x}{2\pi(1-\nu)}\, y \ln r,\nonumber\\ 
&\tl g=\frac{\mu \Omega_x\nu}{\pi(1-\nu)}\, z \ln r,
\end{align}
may be used to reproduce deWit's expressions for the stress and strain fields
of the twist disclination. 
Substituting~(\ref{f_twist_cl1}) and (\ref{stress-ansatz-tw1}) into
(\ref{stress-fe}), the following Helmholtz equations to determine the
``modified'' stress functions follow
\begin{align}
\label{f_fe_tw_x}
\Big(1-\kappa^{-2}\Delta\Big)f&=- \frac{\mu \Omega_x}{2\pi(1-\nu)}  \, z x\ln r ,\nonumber\\
\Big(1-\kappa^{-2}\Delta\Big)F&= \frac{\mu \Omega_x}{2\pi(1-\nu)}\, x\ln r ,\nonumber\\
\Big(1-\kappa^{-2}\Delta\Big)g&=\frac{\mu \Omega_x\nu}{\pi(1-\nu)}\, z \ln r.
\end{align}
The solutions of the modified stress functions are given by 
\begin{align}
\label{f-twist-1}
f&=-\frac{\mu \Omega_x}{2\pi(1-\nu)}\, z x \bigg\{\ln r 
+\frac{2}{\kappa^2 r^2}\Big(1-\kappa r K_1(\kappa r)\Big)\bigg\},\nonumber\\ 
F&=\frac{\mu \Omega_x}{2\pi(1-\nu)}\, y \bigg\{\ln r 
+\frac{2}{\kappa^2 r^2}\Big(1-\kappa r K_1(\kappa r)\Big)\bigg\},\nonumber\\ 
g&=\frac{\mu \Omega_x\nu}{\pi(1-\nu)}\, z\Big\{\ln r +K_0(\kappa r)\Big\},
\end{align}
By means of~(\ref{stress-ansatz-tw1}) and (\ref{f-twist-1}) we
are able to calculate the stress
\begin{align}
\label{T-twist-x}
&\sigma_{xx}=-\frac{\mu \Omega_x}{2\pi(1-\nu)}\, 
\frac{zx}{r^4}\bigg\{\big(x^2-y^2\big)-\frac{4}{\kappa^2r^2}\big(x^2-3y^2\big)
-2 y^2\kappa r K_1(\kappa r)+2\big(x^2-3y^2\big) K_2(\kappa r)\bigg\},\nonumber\\
&\sigma_{yy}=-\frac{\mu \Omega_x}{2\pi(1-\nu)}\, 
\frac{zx}{r^4}\bigg\{\big(x^2+3y^2\big)+\frac{4}{\kappa^2r^2}\big(x^2-3y^2\big)
-2 x^2\kappa r K_1(\kappa r)-2\big(x^2-3y^2\big) K_2(\kappa r)\bigg\},\nonumber\\
&\sigma_{xy}=-\frac{\mu \Omega_x}{2\pi(1-\nu)}\, 
\frac{zy}{r^4}\bigg\{\big(x^2-y^2\big)-\frac{4}{\kappa^2r^2}\big(3x^2-y^2\big)
+2 x^2\kappa r K_1(\kappa r)+2\big(3x^2-y^2\big) K_2(\kappa r)\bigg\},\nonumber\\
&\sigma_{zz}=-\frac{\mu \Omega_x\nu }{\pi(1-\nu)}\, 
\frac{zx}{r^2}\Big\{1-\kappa r K_1(\kappa r)\Big\},\nonumber\\
&\sigma_{zx}=-\frac{\mu \Omega_x}{2\pi(1-\nu)}\, 
\bigg\{(1-2\nu)\big(\ln r+K_0(\kappa r)\big)+\frac{y^2}{r^2}
+\frac{\big(x^2-y^2\big)}{\kappa^2 r^4} \Big(2-\kappa^2 r^2 K_2(\kappa r)\Big)\bigg\},\nonumber\\
&\sigma_{zy}=\frac{\mu \Omega_x}{2\pi(1-\nu)}\, 
\frac{xy}{r^2}\bigg\{1-\frac{2}{\kappa^2 r^2} \Big(2-\kappa^2 r^2 K_2(\kappa r)\Big)\bigg\}.
\end{align}
The trace of the stress tensor reads in this case
\begin{align}
\label{}
\sigma_{kk}=-\frac{\mu \Omega_x(1+\nu)}{\pi(1-\nu)}\, 
\frac{zx}{r^2}\Big\{1-\kappa  r K_1(\kappa r)\Big\}.
\end{align}
The stress~(\ref{T-twist-x}) has its
extreme values in the $xy$-plane:
$|\sigma_{yy}(x,0)|\simeq 0.546\kappa \frac{\mu \Omega_x z}{2\pi(1-\nu)}$ at 
$|x|\simeq 0.996 \kappa^{-1}$,
$|\sigma_{xx}(x,0)|\simeq 0.260 \kappa\frac{\mu \Omega_x z}{2\pi(1-\nu)}$ at 
$|x|\simeq 1.494 \kappa^{-1}$,
$|\sigma_{xy}(0,y)|\simeq 0.260 \kappa\frac{\mu \Omega_x z}{2\pi(1-\nu)}$ at 
$|y|\simeq 1.494 \kappa^{-1}$,
and
$|\sigma_{zz}(x,0)|\simeq 0.399\kappa \frac{\mu \Omega_x z \nu}{\pi(1-\nu)}$ at 
$|x|\simeq 1.114 \kappa^{-1}$.
The stress $\sigma_{zx}$ has at $r=0$ the value:
$\sigma_{zx}(0)\simeq \frac{\mu \Omega_x}{2\pi(1-\nu)}[(1-2\nu)(\gamma
+\ln\frac{\kappa}{2})-0.5]$ and with $\nu=0.3$:
$\sigma_{zx}(0)\simeq \frac{\mu \Omega_x}{2\pi(1-\nu)}[0.4\ln\kappa-0.546]$.

The corresponding elastic strain is given by
\begin{align}
\label{E-twist-x}
&E_{xx}=-\frac{\Omega_x}{4\pi(1-\nu)}\, 
\frac{zx}{r^2}
\bigg\{(1-2\nu)-\frac{2y^2}{r^2}-\frac{4}{\kappa^2r^4}\big(x^2-3y^2\big)\nonumber\\
&\hspace{5cm}-2\left(\frac{y^2}{r^2}-\nu\right)\kappa r K_1(\kappa r)
+\frac{2}{r^2}\big(x^2-3y^2\big) K_2(\kappa r)\bigg\},\nonumber\\
&E_{yy}=-\frac{\Omega_x}{4\pi(1-\nu)}\, 
\frac{zx}{r^2}
\bigg\{(1-2\nu)+\frac{2y^2}{r^2}+\frac{4}{\kappa^2r^4}\big(x^2-3y^2\big)\nonumber\\
&\hspace{5cm}-2\left(\frac{x^2}{r^2}-\nu\right)\kappa r K_1(\kappa r)
-\frac{2}{r^2}\big(x^2-3y^2\big) K_2(\kappa r)\bigg\},\nonumber\\
&E_{xy}=\frac{\Omega_x}{4\pi(1-\nu)}\, 
\frac{zy}{r^2}
\bigg\{1-\frac{2x^2}{r^2}+\frac{4}{\kappa^2r^4}\big(3x^2-y^2\big)
-\frac{2x^2}{r^2}\,\kappa r K_1(\kappa r)
-\frac{2}{r^2}\big(3x^2-y^2\big) K_2(\kappa r)\bigg\},\nonumber\\
&E_{zx}=-\frac{\Omega_x}{4\pi(1-\nu)}\, 
\bigg\{(1-2\nu)\big(\ln r+K_0(\kappa r)\big)+\frac{y^2}{r^2}
+\frac{\big(x^2-y^2\big)}{\kappa^2 r^4} \Big(2-\kappa^2 r^2 K_2(\kappa r)\Big)\bigg\},\nonumber\\
&E_{zy}=\frac{\Omega_x}{4\pi(1-\nu)}\, 
\frac{xy}{r^2}\bigg\{1-\frac{2}{\kappa^2 r^2} \Big(2-\kappa^2 r^2 K_2(\kappa r)\Big)\bigg\}.
\end{align}
The dilatation  reads
\begin{align}
E_{kk}=-\frac{\Omega_x(1-2\nu)}{2\pi(1-\nu)}\, 
\frac{zx}{r^2}\Big\{1-\kappa  r K_1(\kappa r)\Big\}.
\end{align}
The strain~(\ref{E-twist-x}) has the
extreme values ($\nu=0.3$):
$|E_{yy}(x,0)|\simeq 0.308\kappa \frac{ \Omega_x z}{4\pi(1-\nu)}$ at 
$|x|\simeq 0.922 \kappa^{-1}$,
$|E_{xx}(x,0)|\simeq 0.010 \kappa\frac{\Omega_x z}{4\pi(1-\nu)}$ at 
$|x|\simeq 0.218 \kappa^{-1}$, 
$|E_{xx}(x,0)|\simeq 0.054 \kappa\frac{ \Omega_x z}{4\pi(1-\nu)}$ at 
$|x|\simeq 4.130 \kappa^{-1}$, 
and
$|E_{xy}(0,y)|\simeq 0.260 \kappa\frac{\Omega_x z}{4\pi(1-\nu)}$ at 
$|y|\simeq 1.494 \kappa^{-1}$. 
Again, it is interesting to note that 
$E_{xx}(x,0)$ is much smaller than $E_{yy}(x,0)$ within
the core region.
The strain $E_{zx}$ has at $r=0$ the value:
$E_{zy}(0)\simeq \frac{\mu \Omega_x}{4\pi(1-\nu)}[(1-2\nu)(\gamma
+\ln\frac{\kappa}{2})-0.5]$.
The dilatation has its extremum
$|E_{kk}(x,0)|\simeq 0.399\kappa \frac{\Omega_x(1-2\nu)z}{2\pi(1-\nu)}$ at 
$|x|\simeq 1.114 \kappa^{-1}$.

It is interesting to note that there is a relation between the stress and
strain fields of twist disclinations and edge disclinations.
If one replaces $\Omega_x z$ by $-b_y$ in 
the components of the stress $\sigma_{xx}$ -- $\sigma_{zz}$ in~(\ref{T-twist-x})
and in the strain $E_{xx}$ -- $E_{xy}$ in~(\ref{E-twist-x}), the 
stress~(\ref{T-edge-y}) and the strain~(\ref{E-edge-y}) are obtained.

\subsection{Twist disclination: $\frank=(0,\Omega_y,0)$}
For the twist disclination with Frank vector~$\Omega_y$ 
we make a similar ansatz like the ansatz~(\ref{stress-ansatz-tw1}) 
of the twist disclination with  Frank vector~$\Omega_x$. It is given by
\begin{align}
\label{stress-ansatz-tw2}
\tl\sigma {}_{ij}=
\left(\begin{array}{ccc}
\ \,  \pd^2_{yy}\tl f & -\pd^2_{xy}\tl f & -\pd_y \tl F\\
-\pd^2_{xy}\tl f & \pd^2_{xx}\tl f & \pd_x \tl F +\pd_z \tl g\\
-\pd_y \tl F& \pd_x\tl F+\pd_z\tl g&\tl p
\end{array}\right),\quad
\sigma_{ij}=
\left(\begin{array}{ccc}
\ \,  \pd^2_{yy}f & -\pd^2_{xy}f & -\pd_y F\\
-\pd^2_{xy}f & \pd^2_{xx}f & \pd_x F +\pd_z g\\
-\pd_y F& \pd_x F+\pd_z g& p
\end{array}\right),
\end{align}
with the relations
\begin{align}
\tl p=\nu\Delta \tl f=-\pd_y \tl g,\qquad
p=\nu\Delta f=-\pd_y g,
\end{align}
which follow again from $\pd_i\tl\sigma_{zi}=0$ and $\pd_i\sigma_{zi}=0$.
The only one difference between~(\ref{stress-ansatz-tw1}) and 
(\ref{stress-ansatz-tw2}) is the position of the stress function $g$.
In the present case, the ``classical'' stress functions are given by 
\begin{align}
\label{f_twist_cl2}
&\tl f=-\frac{\mu \Omega_y}{2\pi(1-\nu)}\, z y \ln r,\nonumber\\
&\tl F=-\frac{\mu \Omega_y}{2\pi(1-\nu)}\, x \ln r,\nonumber\\ 
&\tl g=\frac{\mu \Omega_y\nu}{\pi(1-\nu)}\, z \ln r.
\end{align}
These three stress functions reproduce the stress and strain fields given by
deWit~\cite{deWit73b}.
If we substitute~(\ref{stress-ansatz-tw2}) and (\ref{f_twist_cl2}) into
(\ref{stress-fe}), we get the following inhomogeneous Helmholtz equations  
\begin{align}
\label{f_fe_tw_y}
\Big(1-\kappa^{-2}\Delta\Big)f&=- \frac{\mu \Omega_yx}{2\pi(1-\nu)}  \, z y\ln r ,\nonumber\\
\Big(1-\kappa^{-2}\Delta\Big)F&=- \frac{\mu \Omega_y}{2\pi(1-\nu)}\, x\ln r ,\nonumber\\
\Big(1-\kappa^{-2}\Delta\Big)g&=\frac{\mu \Omega_y\nu}{\pi(1-\nu)}\, z \ln r.
\end{align}
The solutions of the ``modified'' stress functions reads now 
(see also~\cite{Lazar03c})
\begin{align}
\label{f-twist-y}
f&=-\frac{\mu \Omega_y}{2\pi(1-\nu)}\, z y \bigg\{\ln r 
+\frac{2}{\kappa^2 r^2}\Big(1-\kappa r K_1(\kappa r)\Big)\bigg\},\nonumber\\ 
F&=-\frac{\mu \Omega_y}{2\pi(1-\nu)}\, x \bigg\{\ln r 
+\frac{2}{\kappa^2 r^2}\Big(1-\kappa r K_1(\kappa r)\Big)\bigg\},\nonumber\\ 
g&=\frac{\mu \Omega_y\nu}{\pi(1-\nu)}\, z\Big\{\ln r +K_0(\kappa r)\Big\}.
\end{align}
So we find for the elastic stress of this disclination~\cite{Lazar03c}
\begin{align}
\label{T-twist-y}
&\sigma_{xx}=-\frac{\mu \Omega_y}{2\pi(1-\nu)}\, 
\frac{zy}{r^4}\bigg\{\big(y^2+3x^2\big)+\frac{4}{\kappa^2r^2}\big(y^2-3x^2\big)
-2 y^2\kappa r K_1(\kappa r)-2\big(y^2-3x^2\big) K_2(\kappa r)\bigg\},\nonumber\\
&\sigma_{yy}=-\frac{\mu \Omega_y}{2\pi(1-\nu)}\, 
\frac{zy}{r^4}\bigg\{\big(y^2-x^2\big)-\frac{4}{\kappa^2r^2}\big(y^2-3x^2\big)
-2 x^2\kappa r K_1(\kappa r)+2\big(y^2-3x^2\big) K_2(\kappa r)\bigg\},\nonumber\\
&\sigma_{xy}=\frac{\mu \Omega_y}{2\pi(1-\nu)}\, 
\frac{zx}{r^4}\bigg\{\big(x^2-y^2\big)-\frac{4}{\kappa^2r^2}\big(x^2-3y^2\big)
-2 y^2\kappa r K_1(\kappa r)+2\big(x^2-3y^2\big) K_2(\kappa r)\bigg\},\nonumber\\
&\sigma_{zz}=-\frac{\mu \Omega_y\nu }{\pi(1-\nu)}\, 
\frac{zy}{r^2}\Big\{1-\kappa r K_1(\kappa r)\Big\},\nonumber\\
&\sigma_{zx}=\frac{\mu \Omega_y}{2\pi(1-\nu)}\, 
\frac{xy}{r^2}\bigg\{1-\frac{2}{\kappa^2 r^2} \Big(2-\kappa^2 r^2 K_2(\kappa r)\Big)\bigg\},\nonumber\\
&\sigma_{zy}=-\frac{\mu \Omega_y}{2\pi(1-\nu)}\, 
\bigg\{(1-2\nu)\big(\ln r+K_0(\kappa r)\big)+\frac{x^2}{r^2}
-\frac{\big(x^2-y^2\big)}{\kappa^2 r^4} \Big(2-\kappa^2 r^2 K_2(\kappa r)\Big)\bigg\},
\end{align}
and its trace is given by
\begin{align}
\label{}
\sigma_{kk}=-\frac{\mu \Omega_y(1+\nu)}{\pi(1-\nu)}\, 
\frac{zy}{r^2}\Big\{1-\kappa  r K_1(\kappa r)\Big\}.
\end{align}
\begin{figure}[p]\unitlength1cm
\centerline{
\epsfig{figure=Txx-x.eps,width=7.0cm}
\put(-1.0,1.0){$\kappa y$}
\put(-6.0,1.0){$\kappa x$}
\put(-6.2,-0.3){$\text{(a)}$}
\hspace*{0.4cm}
\put(0,-0.3){$\text{(b)}$}
\put(6.0,1.0){$\kappa y$}
\put(1.0,1.0){$\kappa x$}
\epsfig{figure=Txy-x.eps,width=7.0cm}}
\vspace*{0.2cm}
\centerline{
\epsfig{figure=Tyy-x.eps,width=7.0cm}
\put(-1.0,1.0){$\kappa y$}
\put(-6.0,1.0){$\kappa x$}
\put(-6.0,-0.3){$\text{(c)}$}
\hspace*{0.4cm}
\put(6.0,1.0){$\kappa y$}
\put(1.0,1.0){$\kappa x$}
\put(0,-0.3){$\text{(d)}$}
\epsfig{figure=Tzz-x.eps,width=7.0cm}
}
\vspace*{0.2cm}
\centerline{
\epsfig{figure=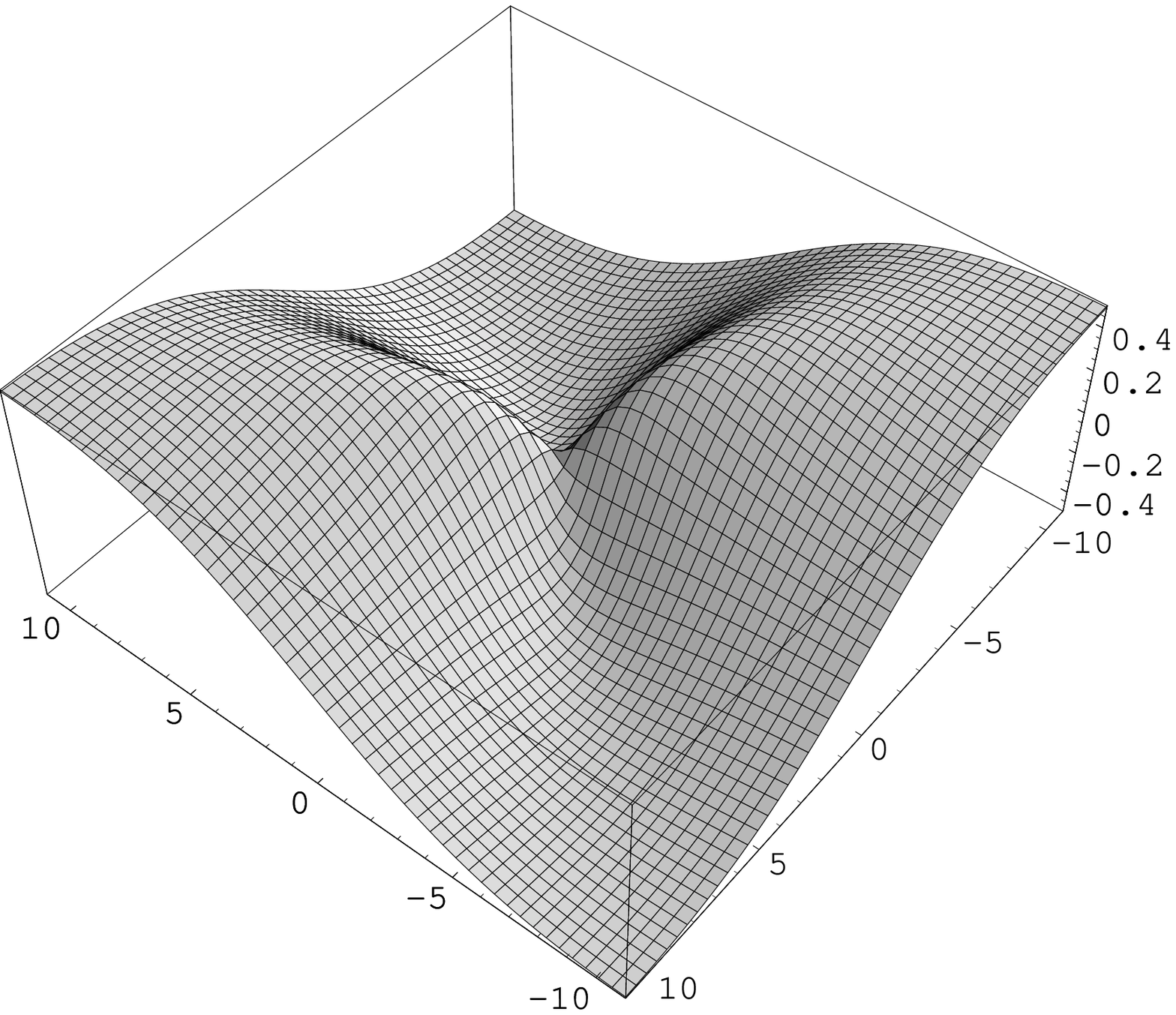,width=7.0cm}
\put(-1.0,1.0){$\kappa y$}
\put(-6.0,1.0){$\kappa x$}
\put(-6.0,-0.3){$\text{(e)}$}
\hspace*{0.4cm}
\put(6.0,1.0){$\kappa y$}
\put(1.0,1.0){$\kappa x$}
\put(0,-0.3){$\text{(f)}$}
\epsfig{figure=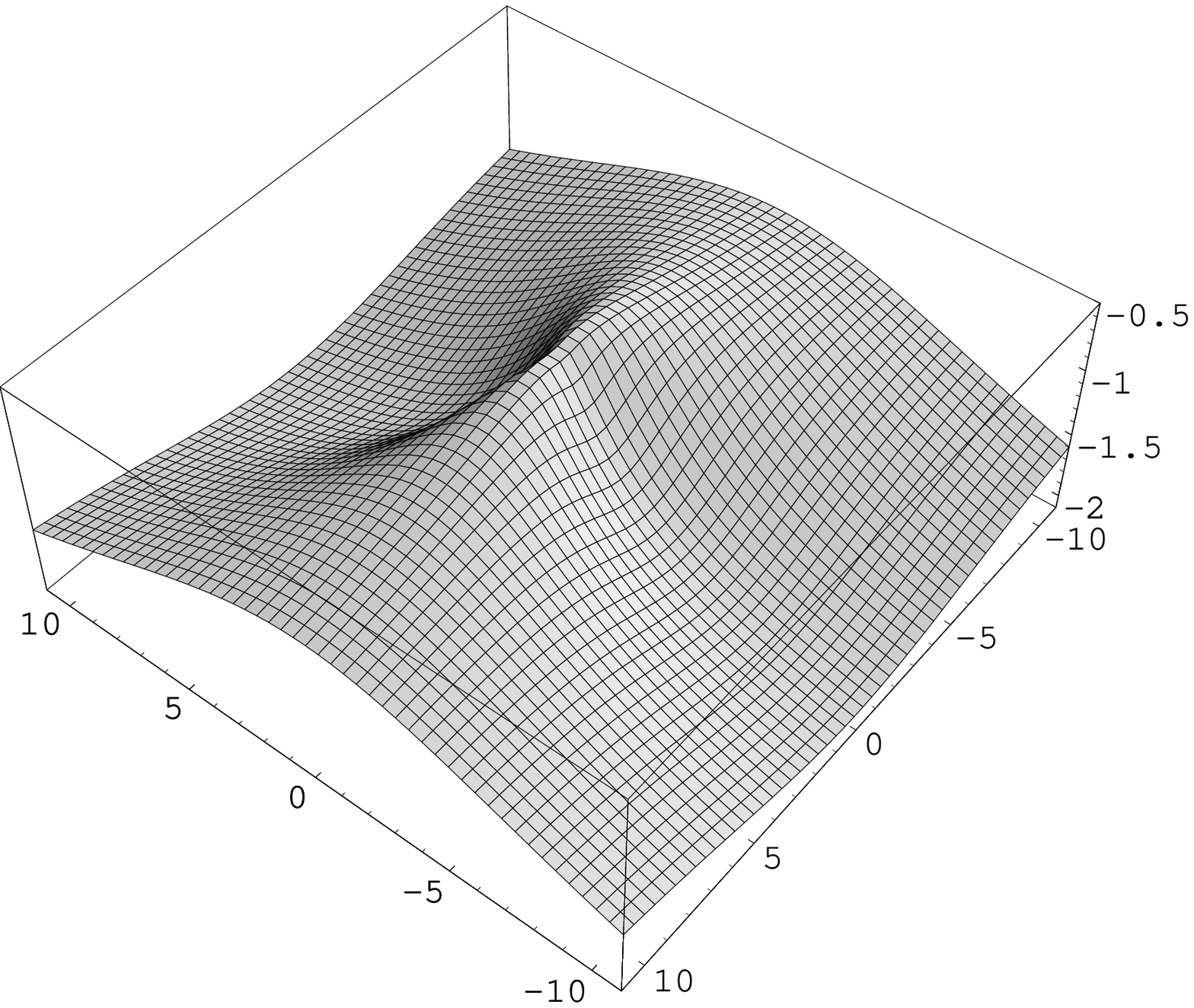,width=7.0cm}}
\caption{Stress of a twist disclination:
(a) $\sigma_{xx}$,
(b) $\sigma_{xy}$,
(c) $\sigma_{yy}$
are given in units of $\mu \Omega_y \kappa z/[2\pi(1-\nu)]$, 
(d) $\sigma_{zz}$
is given in units of $\mu \Omega_y\nu \kappa z/[\pi(1-\nu)]$, 
(e) $\sigma_{zx}$ and
(f) $\sigma_{zy}$
are given in units of $\mu \Omega_y/[2\pi(1-\nu)]$ 
.}
\label{fig:stress-twist-y}
\end{figure}
It can be seen that the stresses have the
following extreme values in the $xy$-plane:
$|\sigma_{xx}(0,y)|\simeq 0.546\kappa \frac{\mu \Omega_y z}{2\pi(1-\nu)}$ at 
$|y|\simeq 0.996 \kappa^{-1}$,
$|\sigma_{yy}(0,y)|\simeq 0.260 \kappa\frac{\mu \Omega_y z}{2\pi(1-\nu)}$ at 
$|y|\simeq 1.494 \kappa^{-1}$,
$|\sigma_{xy}(x,0)|\simeq 0.260 \kappa\frac{\mu \Omega_y z}{2\pi(1-\nu)}$ at 
$|x|\simeq 1.494 \kappa^{-1}$,
$|\sigma_{zz}(0,y)|\simeq 0.399\kappa \frac{\mu \Omega_y\nu z}{\pi(1-\nu)}$ at 
$|y|\simeq 1.114 \kappa^{-1}$ and
$|\sigma_{kk}(0,y)|\simeq 0.399\kappa \frac{\mu \Omega_y(1+\nu)z}{\pi(1-\nu)}$ at 
$|y|\simeq 1.114 \kappa^{-1}$.
The stresses $\sigma_{xx}$,  $\sigma_{yy}$ and  $\sigma_{xy}$ are modified 
near the disclination core ($0\le r\le 12\kappa^{-1}$). 
The stress $\sigma_{zz}$ and the trace $\sigma_{kk}$ are modified
in the region: $0\le r\le 6\kappa^{-1}$.
Far from the disclination line ($r\gg 12\kappa^{-1}$) the modified
and the classical solutions of the stress of a twist disclination coincide.
In addition, it can be seen that at $z=0$ the stresses~$\sigma_{xx}$ -- $\sigma_{zz}$
are zero. 
The stress $\sigma_{zy}$ has at $r=0$ the value:
$\sigma_{zy}(0)\simeq \frac{\mu \Omega_y}{2\pi(1-\nu)}[(1-2\nu)(\gamma
+\ln\frac{\kappa}{2})-0.5]$ and with $\nu=0.3$:
$\sigma_{zy}(0)\simeq \frac{\mu \Omega_y}{2\pi(1-\nu)}[0.4\ln\kappa-0.546]$.
The stress is plotted in Fig.~\ref{fig:stress-twist-y}.

For the elastic strain we obtain
\begin{align}
\label{E-twist-y}
&E_{xx}=-\frac{\Omega_y}{4\pi(1-\nu)}\, 
\frac{zy}{r^2}
\bigg\{(1-2\nu)+\frac{2x^2}{r^2}+\frac{4}{\kappa^2r^4}\big(y^2-3x^2\big)\nonumber\\
&\hspace{5cm}-2\left(\frac{y^2}{r^2}-\nu\right)\kappa r K_1(\kappa r)
-\frac{2}{r^2}\big(y^2-3x^2\big) K_2(\kappa r)\bigg\},\nonumber\\
&E_{yy}=-\frac{\Omega_y}{4\pi(1-\nu)}\, 
\frac{zy}{r^2}
\bigg\{(1-2\nu)-\frac{2x^2}{r^2}-\frac{4}{\kappa^2r^4}\big(y^2-3x^2\big)\nonumber\\
&\hspace{5cm}-2\left(\frac{x^2}{r^2}-\nu\right)\kappa r K_1(\kappa r)
+\frac{2}{r^2}\big(y^2-3x^2\big) K_2(\kappa r)\bigg\},\nonumber\\
&E_{xy}=\frac{\Omega_y}{4\pi(1-\nu)}\, 
\frac{zx}{r^2}
\bigg\{1-\frac{2y^2}{r^2}-\frac{4}{\kappa^2r^4}\big(x^2-3y^2\big)
-\frac{2y^2}{r^2}\,\kappa r K_1(\kappa r)
+\frac{2}{r^2}\big(x^2-3y^2\big) K_2(\kappa r)\bigg\},\nonumber\\
&E_{zx}=\frac{\Omega_y}{4\pi(1-\nu)}\, 
\frac{xy}{r^2}\bigg\{1-\frac{2}{\kappa^2 r^2} \Big(2-\kappa^2 r^2 K_2(\kappa r)\Big)\bigg\},\nonumber\\
&E_{zy}=-\frac{\Omega_y}{4\pi(1-\nu)}\, 
\bigg\{(1-2\nu)\big(\ln r+K_0(\kappa r)\big)+\frac{x^2}{r^2}
-\frac{\big(x^2-y^2\big)}{\kappa^2 r^4} \Big(2-\kappa^2 r^2 K_2(\kappa r)\Big)\bigg\}.
\end{align}
The dilatation is 
\begin{align}
E_{kk}=-\frac{\Omega_y(1-2\nu)}{2\pi(1-\nu)}\, 
\frac{zy}{r^2}\Big\{1-\kappa  r K_1(\kappa r)\Big\}.
\end{align}
Again, if one replaces $\Omega_y z$ by $b_x$ in 
the components of the stress $\sigma_{xx}$ -- $\sigma_{zz}$ in~(\ref{T-twist-y})
and in the strain $E_{xx}$ -- $E_{xy}$ in~(\ref{E-twist-y}), the 
stress~(\ref{T-edge-x}) and the strain~(\ref{E-edge-x}) is reproduced 
(see also the discussion in~\cite{Lazar03c}).
The components of the strain tensor have in the $xy$-plane the
following extreme values ($\nu=0.3$):
$|E_{xx}(0,y)|\simeq 0.308\kappa \frac{\Omega_y z}{4\pi(1-\nu)}$ at 
$|y|\simeq 0.922 \kappa^{-1}$,
$|E_{yy}(0,y)|\simeq 0.010 \kappa\frac{\Omega_y z}{4\pi(1-\nu)}$ at 
$|y|\simeq 0.218 \kappa^{-1}$, 
$|E_{yy}(0,y)|\simeq 0.054 \kappa\frac{\Omega_y z}{4\pi(1-\nu)}$ at 
$|y|\simeq 4.130 \kappa^{-1}$, 
and
$|E_{xy}(x,0)|\simeq 0.260 \kappa\frac{\Omega_y z}{4\pi(1-\nu)}$ at 
$|x|\simeq 1.494 \kappa^{-1}$. 
It is interesting to note that 
$E_{yy}(0,y)$ is much smaller than $E_{xx}(0,y)$ within
the core region. 
The strain $E_{zy}$ has at $r=0$ the value:
$E_{zy}(0)\simeq \frac{\mu \Omega_y}{4\pi(1-\nu)}[(1-2\nu)(\gamma
+\ln\frac{\kappa}{2})-0.5]$.
The dilatation has its extremum
$|E_{kk}(0,y)|\simeq 0.399\kappa \frac{\Omega_y(1-2\nu)z}{2\pi(1-\nu)}$ at 
$|y|\simeq 1.114 \kappa^{-1}$.

It is interesting to note that   
the solutions of stress and strain fields of disclinations given in this 
section agree with the 
expressions earlier obtained by Gutkin and Aifantis~\cite{GA00,Gutkin00} 
by using the technique of Fourier transformation.

Now some remarks on the double stresses of twist disclinations are in order.
Only the components of double stresses of twist disclinations 
$\tau_{xxz}$, $\tau_{yyz}$, $\tau_{xyz}$, $\tau_{zzz}$, $\tau_{zxx}$,  
$\tau_{zxy}$, $\tau_{zyx}$ and $\tau_{zyy}$ are nonsingular. 
Again, they are similar in the form like the stresses of edge dislocations.
The other components   
$\tau_{xxx}$, $\tau_{yyx}$, $\tau_{xyx}$, $\tau_{zzx}$,
$\tau_{xxy}$, $\tau_{yyy}$, $\tau_{xyy}$ and $\tau_{zzy}$ are 
singular at $r=0$ like the double stresses of edge dislocations. 
Moreover these components are zero at $z=0$.

\section{Conclusions}
We investigated two special versions of first gradient elasticity.
One special version of Mindlin's gradient theory has been used in the consideration
of dislocations and disclinations. This theory is a strain gradient
as well as a stress gradient theory. 
We have applied this theory to all three types of dislocations and disclinations.
Using the stress function method, we found modified stress functions for 
dislocations and disclinations. These stress functions are modifications
of the classical ones (e.g. Prandtl's and Airy's stress functions). 
All modified stress functions satisfy two-dimensional 
inhomogeneous Helmholtz equations in which
the classical stress functions are the inhomogeneous parts.  
Using these modified stress functions, exact analytical solutions for 
the elastic stress (Cauchy stress) and the elastic strain fields of all
six types of Volterra defects have been found. 
These stress and strain fields fulfill inhomogeneous Helmholtz equations 
in which the inhomogeneous parts are given by the classical singular 
stress and strain fields, respectively.
The main feature of these solutions is that the unphysical singularities 
at the defect line are eliminated. Thus, the improved stress and strain
fields have no singularities in the core region unlike the classical 
solutions of defects in elasticity which are singular in this region.
The stresses and strains are either zero
or have a finite maximum value at the defect line. The maximum value 
of stresses may serve as a measure of the critical stress level when 
fracture and failure may occur.
In addition, if we equate the maximum shear stress to the cohesive 
shear stress, one can obtain conditions to produce a dislocation or 
disclination of single atomic distance.
In gradient elasticity the maximum value of stresses depend on the gradient
coefficient $\kappa$. Thus, one could test the value of $\kappa$ if
one uses the theoretical shear stresses based on lattice
dynamics calculations. 

The gradient theory considered in this paper contains double stresses 
(hyperstresses) which
are simple gradients of the Cauchy stress. In the case of dislocations 
these double stresses are still singular at $r=0$.
Only for the wedge disclination the double stresses are nonsingular.
For the twist disclinations some components of the double stress tensor 
are nonsingular and the other components have singularities at $r=0$. 

Finally, we notice that the stress fields of dislocations and disclinations
are also solutions in Eringen's nonlocal elasticity by using the nonlocal 
kernel~(\ref{green}) and the condition~(\ref{EC-C}). 
In fact, the nonsingular stresses correspond to the nonlocal ones.  
Eventually, the nonsingular stresses of dislocations and disclinations
can be calculated by means of the convolution of the singular classical
stresses with a nonlocal kernel which is a kind of a distribution
function. It is the convolution with a suitable kernel 
that provides smoothing of the 
``classical'' stress singularities and produces nonsingular
stresses which are the main features in nonlocal elasticity.

\subsection*{Acknowledgement}
This work was supported by the European Network TMR 98-0229
and, in part, by
the European Network RTN ``DEFINO''
with contract number HPRN-CT-2002-00198.
M.L. is grateful to Prof. Elias C. Aifantis and Dr. Mikhail Gutkin  
for helpful discussions and remarks about gradient elasticity.

\begin{appendix}
\renewcommand{\theequation}{\thesection.\arabic{equation}}
\setcounter{equation}{0}
\section{Expansion of Bessel functions}
\label{append}
In this appendix, we give the expansion of modified Bessel functions 
which we need to study the near field 
of nonsingular stresses and strains.
The expansion is given by (see, e.g.,~\cite{Leb})
\begin{align}
\label{K0-exp}
&K_0(\kappa r)\approx -\left[\ln\frac{\kappa r}{2}+\gamma\right]
-\left[\ln\frac{\kappa r}{2}-\big(1-\gamma\big)\right]
\frac{\kappa^2 r^2}{4}
+{\cal O}\big(r^4\big),\\
\label{K1-exp}
&K_1(\kappa r)\approx \frac{1}{\kappa r}
+\left[\ln\frac{\kappa r}{2}-\frac{\big(1-2\gamma\big)}{2}\right]
\frac{\kappa r}{2}
+{\cal O}\big(r^3\big),\\
\label{K2-exp}
&K_2(\kappa r)\approx -\frac{1}{2}+\frac{2}{\kappa^2 r^2}
-\left[\ln\frac{\kappa r}{2}-\left(\frac{3}{4}-\gamma\right)\right]
\frac{\kappa^2 r^2}{8}
+{\cal O}\big(r^4\big),
\end{align}
where $\gamma=0.57721566\ldots$ is Euler's constant. 
The first terms in the expansions~(\ref{K0-exp})--(\ref{K2-exp}) eliminate
the classical singularities of the stresses and strains. 
This elimination of singularities is a kind of ``regularization'' of
stresses and strains. 
The other terms are zero at $r=0$. Thus, the second terms in (\ref{K0-exp})--(\ref{K2-exp})
describe the behaviour near the defect line in the first order of the 
expansion. 

\end{appendix}


\begin{thebibliography}{99}
\bibitem{Kroener63} E.~Kr{\"o}ner, Int. J. Engng. Sci.~{\bf 1} (1963) 261.
\bibitem{KD66} E.~Kr{\"o}ner and B.K.~Datta, Z.~Phys.~{\bf 196} (1966) 203.
\bibitem{Mindlin64} R.D.~Mindlin, Arch. Rat. Mech. Anal.~{\bf 16} (1964) 51.
\bibitem{Mindlin65} R.D.~Mindlin, Int. J. Solids Struct.~{\bf 1} (1965) 417.
\bibitem{ME68} R.D.~Mindlin and N.N.~Eshel, Int. J. Solids Struct.~{\bf 4} (1968) 109.
\bibitem{GR64a} A.E.~Green and R.S.~Rivlin, Arch. Rat. Mech. Anal.~{\bf 16} (1964) 325. 
\bibitem{GR64b} A.E.~Green and R.S.~Rivlin, Arch. Rat. Mech. Anal.~{\bf 17} (1964) 113. 
\bibitem{Toupin64} R.A.~Toupin, Arch. Rat. Mech. Anal.~{\bf 17} (1964) 85. 
\bibitem{Germain} P.~Germain, SIAM J. Appl. Math.~{\bf 25} (1973) 556.
\bibitem{Wu92} C.H.~Wu, Quart. Appl. Mat.~{\bf 50} (1992) 73. 
\bibitem{Maugin93} G.A.~Maugin, {\it Material Inhomogeneities in Elasticity}, 
	Chapman and Hall, London (1993).
\bibitem{hehl76} F.W.~Hehl, G.D.~Kerlick and P. von der Heyde, 
	Z. Naturforsch.~{\bf 31 a} (1976) 111.
\bibitem{hehl97} F.~Gronwald and F.W.~Hehl, {\it Stress and hyperstress as fundamental
        concepts in continuum mechanics and in relativistic field theory},
        in {\it Advances in Modern Continuum Dynamics}, International Conference 
	in Memory of Antonio Signorini, Isola d'Elba, June 1991, G. Ferrarese, 
	ed. Pitagora Editrice, Bologna (1993) pp. 1-32; 
	Eprint Archive {\tt http://www.arXiv.org/abs/gr-qc/9701054}.
\bibitem{deWit73b} R.~deWit, 
        J.~Res.~Nat. Bur. Stand. (U.S.)~{\bf 77A} (1973) 607.
\bibitem{EL88} D.G.B.~Edelen and D.C.~Lagoudas, {\it Gauge theory and defects in 
	solids}, in: {\it Mechanics and Physics of Discrete System}, Vol.~1,
	G.C.~Sih, ed., North-Holland, Amsterdam (1988).
\bibitem{Edelen96} D.G.B.~Edelen, 
        Int. J. Engng. Sci.~{\bf 34} (1996) 81.
\bibitem{Lazar02a} M.~Lazar, 
        J. Phys. A: Math. Gen.~{\bf 35} (2002) 1983.
\bibitem{Lazar02b} M.~Lazar, 
        Ann. Phys.~(Leipzig)~{\bf 11} (2002) 635.
\bibitem{Lazar03a} M.~Lazar, 
        J.~Phys.~A: Math. Gen.~{\bf 36} (2003) 1415.
\bibitem{Feynman} R.P.~Feynman, {\it Lectures on Gravitation}, 
	Lecture notes by F.B.~Morinigo and W.G.~Wagner 
	(California Institutes of Technology, Pasadena, California 1962/63),
	Addison-Wesley (1995).
\bibitem{KV92} M.O.~Katanaev and I.V.~Volovich, 
	Ann.~Phys.~(N.Y.)~{\bf 216} (1992) 1.
\bibitem{Kroener58} E.~Kr{\"o}ner, {\it Kontinuumstheorie der Versetzungen und Eigenspannungen},
	Erg.~Angew. Math.~{\bf 5} (1958), 1-179.
\bibitem{Kroener81} E.~Kr{\"o}ner, {\it Continuum Theory of Defects}, in:
        {\it Physics of Defects} (Les Houches, Session 35), R.~Balian et al., eds.,
        North-Holland, Amsterdam (1981) p.~215.
\bibitem{Lazar00} M.~Lazar, 
  	Ann. Phys.~(Leipzig)~{\bf 9} (2000) 461.
\bibitem{Malyshev00} C.~Malyshev, 
         Ann.~Phys. (N.Y.)~{\bf 286} (2000) 249.
\bibitem{Ex98} G.~Exadaktylos, Int. J. Solids Structures~{\bf 35} (1998) 421.
\bibitem{Fann} A.C.~Fannjiang, Y.-S.~Chan and G.H.~Paulino, SIAM J. Appl. Math.~{\bf 62} (2002) 1066.
\bibitem{Georg} H.G.~Georgiadis, 
	ASME J. Appl. Mech.~{\bf 70} (2003) 517.
\bibitem{AA97} B.S.~Altan and E.C.~Aifantis, J.~Mech. Behav. Mater.~{\bf 8} 
        (1997) 231.
\bibitem{GA99b} M.Yu.~Gutkin and E.C.~Aifantis, 
        Phys. Stat. Sol. (b)~{\bf 214} (1999) 245.
\bibitem{RA93} C.Q.~Ru and E.C.~Aifantis, 
	{\it Some studies on boundary value problems in gradient elasticity},
	(1993), unpublished.
\bibitem{GA99} M.Yu.~Gutkin and E.C.~Aifantis, 
        Scripta Mater.~{\bf 40} (1999) 559.
\bibitem{GA00} M.Yu.~Gutkin and E.C.~Aifantis, in:
        {\it Nanostructured Film and Coatings},
        NATO ARW Series, High Technology, Vol. 78, ed. by G.M.~Chow et al. 
        (Kluwer, Dodrecht, 2000) p.~247.
\bibitem{Gutkin00} M.Yu.~Gutkin, Rev. Adv. Mater. Sci.~{\bf 1} (2000) 27.
\bibitem{Eringen83} A.C.~Eringen, J. Appl. Phys.~{\bf 54} (1983) 4703.
\bibitem{Eringen85} A.C.~Eringen, {\it Nonlocal Continuum Theory for Dislocations
        and Fracture}, in: {\it The Mechanics of Dislocations},
        Eds. E.C.~Aifantis and J.P.~Hirth, American Society of Metals, 
        Metals Park, Ohio (1985) p.~101.
\bibitem{Eringen90} A.C.~Eringen, {\it On Screw Dislocations and Yield}, in:
        Elasticity, Mathematical Methods and Applications, G.G.~Eason and 
        R.W.~Odgen, eds., Ellis Harwood, Chichester (1990) p.~87.
\bibitem{Eringen02} A.C.~Eringen, {\it Nonlocal Continuum Field Theories},
        Springer, New York (2002).
\bibitem{AE83} N.~Ari and A.C.~Eringen, Cryst. Lattice Defects Amorph. Mat.~{\bf 10} 
        (1983) 33.
\bibitem{Aifan92} E.C.~Aifantis, Int. J. Engng. Sci.~{\bf 30} (1992) 1279.
\bibitem{RA} C.Q.~Ru and E.C.~Aifantis, 
	Acta Mech.~{\bf 101} (1993) 59 .
\bibitem{Aifantis03} E.C.~Aifantis, Mechanics of Materials~{\bf 35} (2003) 259.
\bibitem{GA96} M.Yu.~Gutkin and E.C.~Aifantis, 
        Scripta Mater.~{\bf 35} (1996) 1353.
\bibitem{GA97} M.Yu.~Gutkin and E.C.~Aifantis, 
        Scripta Mater.~{\bf 36} (1997) 129.
\bibitem{GVV04} H.G.~Georgiadis, I.~Vardoulakis and E.G. Velgaki,
	J. Elasticity~{\bf 74} (2004) 17.
\bibitem{Lazar02d} M.~Lazar, 
        Comput. Mater. Sci.~{\bf 28} (2003) 419.
\bibitem{LMA04} 
        M.~Lazar, G.A.~Maugin and E.C.~Aifantis, 
	{\it Dislocations in second strain gradient elasticity}, (2004), 
	submitted. 
\bibitem{RV} A.E.~Romanov and V.I.~Vladimirov, {\it Disclinations in crystalline
        solids}, in: {\it Dislocations in Solids Vol.~9}, 
        F.R.N.~Nabarro, ed., North-Holland (1992) p.~191.
\bibitem{Lazar03b} M.~Lazar, 
	Phys. Lett.~A~{\bf 311} (2003) 416. 
\bibitem{Lazar03c} M.~Lazar, 
	J.~Phys.: Condens. Matter~{\bf 15} (2003) 6781.	
\bibitem{Leb} N.N.~Lebedev, {\it Special functions and their applications},
	Dover, New York (1972).
\end{thebibliography}
\end{document}